\documentclass[journal]{IEEEtran}
\ifCLASSINFOpdf
\else
   \usepackage{graphicx}
   \graphicspath{{../eps/}}
   \DeclareGraphicsExtensions{.eps}
\fi
\usepackage{fancyhdr}
\usepackage{amsmath,amssymb,amstext}
\usepackage{subeqnarray}
\usepackage{cite}
\usepackage{hhline}
\usepackage{bm}
\usepackage{soul}
\usepackage{slashbox}
\usepackage{times,amsmath,amsfonts}
\usepackage{subfigure}
\usepackage{amsmath,amssymb,bm}
\usepackage{graphicx,picture}
\usepackage{cite}
\usepackage{subfigure}
\usepackage{subeqnarray}
\usepackage{cases}
\usepackage{color}
\usepackage{overpic}
\usepackage{multirow}
\usepackage{booktabs}
\usepackage{epstopdf}
\usepackage[noend]{algpseudocode}
\usepackage{algorithmicx,algorithm}

\definecolor{r}{rgb}{1,0,0}
\definecolor{b}{rgb}{0,0,1}
\definecolor{k}{rgb}{0,1,1}

\usepackage{upgreek}

\newcounter{saveeqn}%

\allowdisplaybreaks
\DeclareMathSymbol{\Phi}{\mathord}{letters}{8}

\begin{document}
\title{Integrated Sensing and Communication with Delay Alignment Modulation: Performance Analysis and Beamforming Optimization}

\author{
\IEEEauthorblockN{Zhiqiang Xiao,~\IEEEmembership{Student Member, IEEE} and Yong Zeng,~\IEEEmembership{Member, IEEE}}

\thanks{
This work was supported by the National Key R\&D Program of China with grant number 2019YFB1803400,  by the Natural Science Foundation of China under grant 62071114, by the Fundamental Research Funds for the Central Universities of China under grant 3204002004A2 and 2242022k30005.
Part of this work has been presented at the IEEE ICC 2022, Seoul, South Korea, 16-20 May 2022 \cite{xiao2022integrated}.

The authors are with the National Mobile Communications Research Laboratory and Frontiers Science Center for Mobile Information Communication and Security, Southeast University, Nanjing 210096, China. Y. Zeng is also with the Purple Mountain Laboratories, Nanjing 211111, China (e-mail: \{zhiqiang\_xiao, yong\_zeng\}@seu.edu.cn). (\emph{Corresponding author: Yong Zeng.})
}
}
\maketitle

\begin{abstract}
\textsl{\textbf{Delay alignment modulation}} (DAM) has been recently proposed to enable manipulable channel delay spread for efficient single- or multi-carrier communications. In particular, with perfect delay alignment, inter-symbol interference (ISI) can be eliminated even with single-carrier (SC) transmission, without relying on sophisticated channel equalization.
The key ideas of DAM are {\it delay pre-compensation} and {\it path-based beamforming}, so that all multi-path signal components may arrive at the receiver simultaneously and be superimposed constructively, rather than causing the detrimental ISI.
Compared to the classic orthogonal frequency division multiplexing (OFDM) transmission, DAM-enabled SC communication has several appealing advantages, including low peak-to-average-power ratio (PAPR) and high tolerance for Doppler frequency shift, which renders DAM also appealing for radar sensing.
Therefore, in this paper, DAM is investigated for integrated sensing and communication (ISAC) systems.
We first study the output signal-to-noise ratios (SNRs) for ISI-free SC communication and radar sensing, and then derive the closed-form expressions for DAM-based sensing in terms of the ambiguity function (AF) and integrated sidelobe ratio (ISR).
Furthermore, we study the beamforming design problem for DAM-based ISAC to maximize the communication SNR while guaranteeing the sensing performance in terms of the sensing SNR and ISR.
Finally, we provide performance comparison between DAM and OFDM for ISAC, and it is revealed that DAM signal may achieve better communication and sensing performance, thanks to its low PAPR, reduced guard interval overhead, as well as higher tolerance for Doppler frequency shift.
Simulation results are provided to show the great potential of DAM for ISAC.
\end{abstract}


\section{Introduction}
The sixth generation (6G) mobile communication networks are expected to not only further enhance the existing wireless communication services, but also bring in the new sensing capabilities, so as to bridge the cyber and real worlds with intelligence \cite{saad2019vision,dang2020should,you2021towards,tong20216g,xiao2022overview}.
To realize such ambitious visions, integrated sensing and communication (ISAC) has attracted extensive research attentions recently \cite{paul2016survey,feng2020joint,tan2021integrated,liu2022integrated,zhang2021enabling}.
ISAC aims at efficiently utilizing the precious radio resources, hardware, and infrastructure platform to achieve dual functions of sensing and communication.
Furthermore, mutualism between sensing and communication could also be achieved, say via sensing-aided communication and communication-aided sensing enhancement \cite{zheng2019radar,cui2021integrating,wang2021SNR}.

One fundamental problem for ISAC is to find the suitable waveforms to simultaneously guarantee the desired performance for both communication and sensing \cite{zhang2021overview,liu2022survey,xiao2022waveform}.
Most existing approaches on waveform design for ISAC can be classified into two categories, namley, that based on {\it radar-communication coexistence} (RCC) \cite{han2013joint,shi2017power,sodagari2012projection} and {\it dual-function radar-communication} (DFRC) \cite{hassanien2016signaling,sturm2011waveform,kumari2019adaptive}.
For RCC, the radar sensing and communication functions are designed separately and their signals are treated as the detrimental interference by each other.
As a consequence, in order to eliminate the interference, the sensing and communication signals may occupy orthogonal radio resources by time-division \cite{han2013joint}, frequency-division \cite{shi2017power}, or space-division \cite{sodagari2012projection}.
Compared with RCC-based design, the DFRC-based approach is a more aggressive technique towards ISAC, which aims at fully utilizing the radio and hardware resources in a shared manner with  unified waveforms.
The research efforts on DFRC waveforms can be further classified as {\it radar-centric} and {\it communication-centric} designs.
For radar-centric methods, radar sensing is treated as the primary goal and communication symbols are usually embedded into the classic radar waveforms, such as chirp or frequency modulation continuous waveform (FMCW), via radar pulse modulation \cite{nowak2016co}, index modulation \cite{huang2020majorcom}, or beampattern modulation \cite{hassanien2015dual}.
However, such techniques suffer from extremely low communication rate since information embedding is usually applied on the slow-time scale across radar pulses, leading to the communication rate that is limited by the pulse repetition frequency (PRF)~\cite{liu2022survey,xiao2022waveform}.

Different from radar-centric methods, communication-centric methods try to utilize the standard communication waveforms to achieve both sensing and communication purposes.
In particular, orthogonal frequency division multiplexing (OFDM)-based ISAC system has been extensively studied \cite{sturm2011waveform,braun2014ofdm,barneto2019full,keskin2021mimo}.
For communication, OFDM is a mature technology that has gained great success in WiFi, 4G and 5G \cite{heath2018foundations}.
On the other hand, for sensing, OFDM radar can decouple the delay and Doppler frequency estimation by simply applying the Fast Fourier Transform (FFT) and inverse-FFT (IFFT) \cite{sturm2011waveform}.
However, as a multi-carrier transmission technology, OFDM-based ISAC suffers from several practical drawbacks like high peak-to-average-power ratio (PAPR) \cite{han2005overview}, vulnerability to carrier frequency offset (CFO) \cite{sathananthan2001probability}, and Doppler-induced inter-carrier interference (ICI) \cite{wang2006performance}.
In particular, with high PAPR, the transmit power has to be backoff to avoid nonlinear signal distortion, which restricts the maximum communication and sensing range.
Furthermore, in high mobility scenarios, the ICI issue will become a major impairment for communication and sensing performance \cite{wang2006performance,hakobyan2017novel}.
To tackle the above issues, various PAPR reduction or ICI mitigation techniques have been proposed, such as amplitude clipping, tone reservation, and single-carrier frequency-division multiple access (SC-FDMA)~\cite{han2005overview,hakobyan2017novel}.
Recently, a new multi-carrier technique termed {\it orthogonal time frequency space} (OTFS) modulation has been proposed \cite{hadani2017orthogonal}, which modulates data symbols in the delay-Doppler domain, so that each data symbol is spread over the entire time-frequency diversity.
In \cite{raviteja2019orthogonal,gaudio2020effectiveness,yuan2021integrated}, OTFS was studied for radar systems, which showed its superiority than OFDM in high-mobility scenarios.
However, the above techniques usually incur non-negligible performance loss or require rather complicated signal processing.

In this paper, we investigate ISAC with a novel transmission technology, termed {\it delay alignment modulation} (DAM)~\cite{lu2022delay,lu2022delay2}.
By exploiting the high spatial resolution with large antenna arrays and multi-path sparsity of millimeter wave (mmWave) or  Terahertz wireless channels, the key ideas of DAM are {\it delay pre-compensation} and {\it path-based beamforming}.
Specifically, by deliberately introducing symbol delays at the transmitter side to match with the corresponding channel path delays, together with per-path based beamforming, all multi-path signal components may arrive at the receiver simultaneously and be superimposed constructively, rather than causing the detrimental inter-symbol interference (ISI) \cite{lu2022delay}.
As a result, DAM enables ISI-free communication by low complexity single-carrier (SC) transmission without relying on sophisticated channel equalization.
Furthermore, compared to OFDM, DAM can resolve the practical issues like high PAPR, vulnerability to CFO, and ICI.
Such appealing features also render DAM attractive for radar sensing.
Specifically, with reduced PAPR, DAM may allow higher transmit power than OFDM before those nonlinear devices like power amplifiers get saturated.
Moreover, different from OFDM that typically tolerates the Doppler frequency shift only about 10\% of the carrier spacing \cite{raviteja2019orthogonal}, the tolerable Doppler frequency shift is significantly enhanced for DAM, which thus improves the velocity estimation range.
The main contributions of this paper are summarized as follows:

\begin{itemize}
\item First, we present the system model for DAM-based ISAC, which aims to simultaneously achieve ISI-free communication to a user equipment (UE) and sense a radar target.
    The performance analysis of the proposed DAM-ISAC is given, where both the communication and sensing output signal-to-noise ratios (SNRs) are derived.
    Furthermore, the closed-form expression of the ambiguity function (AF) for sensing is analyzed, and the maximum peak sidelobe ratio (PSR) and the integrated sidelobe ratio (ISR) of the AF are derived.

\item Second, for the proposed DAM-ISAC system, we formulate a beamforming optimization problem to maximize the communication SNR while guaranteeing a prescribed  sensing performance in terms of output SNR and ISR.
    The formulated problem is non-convex, for which a semidefinite relaxing (SDR) based method is proposed to obtain an effective solution.

\item Third, we provide a performance comparison between DAM- and OFDM-based ISAC, in terms of the sensing ambiguity and resolution, as well as the PAPR and SNR.
    The results demonstrate that DAM outperforms OFDM in high mobility scenarios, and has low PAPR and smaller guard interval overhead.

\end{itemize}

The rest of this paper is organized as follows. Section~\ref{system model} presents the system model and introduces the key ideas of the DAM.
In Section~\ref{DAM performance}, we first present the signal processing procedures of DAM-ISAC, followed by a comprehensive sensing performance analysis, including the sensing SNR, AF, PSR, and ISR.
Furthermore, a beamforming optimization problem is studied for DAM-ISAC.
In Section~\ref{DAMversusOFDM}, we provide the performance comparison for DAM- and OFDM-based sensing in terms of the sensing ambiguity, resolution, PAPR and SNR.
In Section~\ref{numerical}, numerical results are provided to evaluate the performance of the proposed DAM-ISAC.
Finally, we conclude this paper in Section~\ref{conclude}.

\section{System Model And Delay Alignment Modulation}\label{system model}
\begin{figure} 
  \centering
  \includegraphics[width=0.38\textwidth]{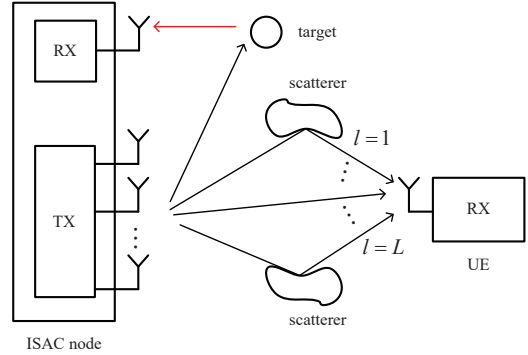}
  \caption{A monostatic ISAC system, where an ISAC node wishes to simultaneously communicate with a UE and sense a target.}\label{system}\vspace{-0.6cm}
\end{figure}
As shown in Fig.~\ref{system}, we consider a monostatic ISAC system, where a multi-antenna ISAC node wishes to simultaneously communicate with a single-antenna  UE and sense a radar target.
The ISAC node is equipped with $M\gg 1$ transmit antennas and one receiving antenna.
We consider a frequency-selective multi-path propagation environment between the ISAC transmitter and the UE.
For one particular channel coherent block, the discrete-time equivalent multiple-input-single-output (MISO) communication channel can be expressed as
\begin{equation}\label{comm_model}
\mathbf{h}_c^H[n] = \sum\nolimits_{l=1}^L\mathbf{h}_l^H\delta[n-n_l],
\end{equation}
where $L$ is the number of temporal-resolvable multi-paths  with delay resolution $T_s=1/B$, with $B$ being the system bandwidth; $\mathbf{h}_l\in\mathbb{C}^{M\times1}$ denotes the complex-valued channel coefficient and $n_l=\mathrm{round}(\tilde{\tau}_l/T_s)$ denotes the normalized delay for the $l$th path, with $\tilde{\tau}_l$ denoting the delay in seconds.
The channel delay spread is $n_{d}=n_{\max}-n_{\min}$, where $n_{\max}=\max\limits_{1\le l\le L}n_l$ and $n_{\min}=\min\limits_{1\le l\le L}n_l$ denote the maximum and minimum delays among the $L$ paths, respectively.

For radar sensing, we assume that there is a clear line-of-sight (LoS) link between the ISAC node and the radar target, and the self-interference of the ISAC node as well as the clutters have been suppressed by applying appropriate cancellation techniques \cite{richards2010principles}.
Note that since the sensing target usually has higher moving speed than the communication UE, which results in higher Doppler frequency shift.
Therefore, different from \eqref{comm_model}, the Doppler of the radar sensing channel is explicitly considered, which is given by
\begin{equation}\label{radar_model}
\mathbf{h}_s^H[n,m] = \alpha\delta[m-\tau]e^{j2\pi\nu nT_s}\mathbf{a}^H(\theta),
\end{equation}
where $\alpha$ denotes the complex-valued channel coefficient including the impact of radar cross section (RCS) of the target;
$\tau=\mathrm{round}(\tilde{\tau}/T_s)$ is the normalized two-way propagation delay, with $\tilde{\tau}$ denoting the delay in second;
$\nu=~\frac{2v}{\lambda}$ is the Doppler frequency caused by the motion of the target with the radial velocity $v$, and $\lambda$ is the carrier wavelength;
$\mathbf{a}(\theta)\in\mathbb{C}^{M\times1}$ is the steering vector of the transmit antenna array, with $\theta$ denoting the direction of the sensing target.
Note that one practical scenario of the considered ISAC system is the cellular-connected unmanned aerial vehicle (UAV) \cite{zeng2018cellular}, where the sensing target corresponds to the UAV, while the ISAC node corresponds to the base station (BS) that wishes to simultaneously communicate with the ground UE and track the UAV in the airspace.

Signals propagated over the time-dispersive multi-path channel in \eqref{comm_model} arrive the receiver via different paths with various delays, which cause the detrimental ISI.
In contemporary communication systems, OFDM is the dominate technology to mitigate ISI, which transforms the frequency-selective channel \eqref{comm_model} into multiple frequency-flat sub-channels with overlapped spectral.
However, OFDM is known to suffer from several issues like high PAPR and sensitive to CFO.
To resolve such issues, a novel transmission technique known as DAM was recently proposed in \cite{lu2022delay,lu2022delay2}.
With DAM-enabled SC transmission, the transmitted signal is
\begin{equation}\label{dam_tx}
\mathbf{x}[n]=\sum\nolimits_{l=1}^{L}\mathbf{f}_ls[n-\kappa_l],
\end{equation}
where $s[n]$ denotes the independent and identically distributed (i.i.d.) information-bearing symbols with normalized power $\mathbb{E}[|s[n]|^2]=1$;
$\kappa_l\geq 0$ is the deliberately introduced delay pre-compensation for the $l$th path, with $\kappa_l\neq\kappa_{l'},\forall l\neq l'$; and $\mathbf{f}_l\in\mathbb{C}^{M\times1}$ denotes per-path based transmit beamforming vector for the $l$th path.
The average transmit power of $\mathbf{x}[n]$ is
\begin{equation}\label{power_const}
\mathbb{E}\left[\left\|\mathbf{x}[n]\right\|^2\right] = \sum_{l=1}^L\mathbb{E}\left[\left\|\mathbf{f}_ls\left[n-\kappa_l\right]\right\|^2\right] = \sum_{l=1}^L\left\|\mathbf{f}_l\right\|^2\le P_t,
\end{equation}
where $P_t$ denotes the average transmit power constraint.
For the communication UE with the channel impulse response given in \eqref{comm_model}, the received signal is
\begin{equation}\label{dam_rx}
\begin{aligned}
y_c[n] &= \mathbf{h}_c^H[n]\ast\mathbf{x}[n] + z[n]\\
& = \sum\limits_{l=1}^{L}\sum\limits_{l'=1}^{L}\mathbf{h}_l^H\mathbf{f}_{l'}s[n-\kappa_{l'}-n_l]+z[n],
\end{aligned}
\end{equation}
where $\ast$ denotes the linear convolution, and $z[n]\sim\mathcal{CN}(0,\sigma^2)$ is the additive white Gaussian noise (AWGN).

By letting $\kappa_{l'}=~n_{\max}-n_{l'}\geq 0, l'=1,\cdots,L$, \eqref{dam_rx} can be written as
\begin{equation}\label{dam_opt}
\begin{aligned}
&y_c[n] = \left(\sum\nolimits_{l=1}^{L}\mathbf{h}_l^H\mathbf{f}_l\right)s[n-n_{\max}]\\
&+\sum\nolimits_{l=1}^{L}\sum\nolimits_{l'\neq l}^{L}\mathbf{h}_l^H\mathbf{f}_{l'}s[n-n_{\max}+n_{l'}-n_l]+z[n],
\end{aligned}
\end{equation}
where all the multi-path components in the first term are aligned with a common delay $n_{\max}$, and they contribute to the desired signal, while the second term causes the ISI.
Fortunately, the detrimental ISI can be effectively suppressed by applying path-based transmit beamforming \cite{lu2022delay}.
In particular, if $\left\{\mathbf{f}_{l'}\right\}_{l'=1}^L$ are designed such that $\mathbf{h}_{l}^H\mathbf{f}_{l'}=0,\forall l'\neq l$, the received DAM signal in \eqref{dam_opt} reduces to
\begin{equation}\label{ISI-ZF}
y_c[n] = \left(\sum\nolimits_{l=1}^{L}\mathbf{h}_l^H\mathbf{f}_l\right)s[n-n_{\max}] + z[n].
\end{equation}
As a result, the original time-dispersive multi-path channel has been transformed into an ISI-free AWGN channel with a single delay $n_{\max}$, and the resulting SNR is
\begin{equation}\label{comm_snr}
\gamma_c = {\left|\sum\nolimits_{l=1}^L\mathbf{h}_l^H\mathbf{f}_l\right|^2}/\sigma^2.
\end{equation}

On the other hand, the DAM signal in \eqref{dam_tx} also has several appealing features for radar sensing.
Specifically, different from the conventional OFDM radar \cite{sturm2011waveform} which suffers from high PAPR issue, the DAM signal in \eqref{dam_tx} utilizes SC transmission, which is expected to have lower PAPR and higher power efficiency.
Moreover, for OFDM, the orthogonality among subcarriers may be destroyed by high Doppler frequency, resulting in ICI and degraded sensing performance \cite{keskin2021mimo}.
Typically, the maximum tolerable Doppler frequency of OFDM is about $10$\% of the subcarrier spacing \cite{raviteja2019orthogonal}.
By contrast, for DAM, since it is a SC transmission scheme, the Doppler frequency only results in the phase variation of the received signal but does not cause interference.
Therefore, DAM is expected to have higher Doppler frequency (or velocity) estimation range.
In the following, we first analyze the sensing performance of DAM, in terms of the sensing output SNR, AF, PSR, and ISR of the DAM signal, and then study the beamforming optimization problem for DAM-ISAC.

\section{Performance Analysis and Beamforming Optimization for DAM-ISAC }\label{DAM performance}
\subsection{DAM for sensing}
For DAM-based sensing, let $N$ be the number of symbol durations for one coherent processing interval (CPI).
The transmitted DAM signal in \eqref{dam_tx} over one CPI can be expressed as
\begin{equation}\label{dam_tx2}
\bar{\mathbf{X}}[n] = \big[\mathbf{x}[n-(N-1)],\cdots,\mathbf{x}[n]\big]\in\mathbb{C}^{M\times N}.
\end{equation}
Denote by $\mathbf{F}\triangleq\left[\mathbf{f}_1,\cdots,\mathbf{f}_L\right]\in\mathbb{C}^{M\times L}$ the transmit beamforming matrix, and $\bar{\mathbf{S}}[n]\triangleq\left[\mathbf{s}[n-\kappa_1],\cdots,\mathbf{s}[n-\kappa_L]\right]^T\in\mathbb{C}^{L\times N}$ the transmitted symbols, where
\begin{equation}
\mathbf{s}[n-\kappa_l]=\big[s[n-(N-1)-\kappa_l],\cdots,s[n-\kappa_l]\big]^T\in\mathbb{C}^{N\times 1}\notag,
\end{equation}
$l=1,\cdots,L$.
Based on \eqref{dam_tx}, the transmitted DAM signal over one CPI can be written as
\begin{equation}\label{dam_tx3}
\bar{\mathbf{X}}[n] = \mathbf{F}\bar{\mathbf{S}}[n].
\end{equation}

For a sensing target with the channel in \eqref{radar_model}, the received echo signal over one CPI is
\begin{equation}\label{dam_echo}
\begin{aligned}
\bar{\mathbf{y}}_s^H[n] &= \alpha\mathbf{a}^H(\theta)\bar{\mathbf{X}}[n-\tau]\mathrm{diag}\left(\mathbf{d}^T[\nu]\right) + \mathbf{z}^H \\
&=\alpha\mathbf{a}^H(\theta)\mathbf{F}\bar{\mathbf{S}}[n-\tau]\mathrm{diag}\left(\mathbf{d}^T[\nu]\right) + \mathbf{z}^H,
\end{aligned}
\end{equation}
where $\mathbf{d}[\nu]\triangleq[e^{j2\pi \nu(n-(N-1))T_s},\cdots,e^{j2\pi \nu nT_s}]^T\in\mathbb{C}^{N\times 1}$ denotes the phase rotation over one CPI caused by the Doppler frequency $\nu$, and $\mathbf{z}\in\mathbb{C}^{N\times 1}$ is the i.i.d. AWGN vector with $\mathbb{E}[\mathbf{z}\mathbf{z}^H]=\sigma^2\mathbf{I}_N$.
Note that for target sensing, the parameters of interest include $\theta$, $\tau$, and $\nu$.
As this paper mainly focuses on the delay-Doppler domain sensing, we assume that the target direction $\theta$ in \eqref{dam_echo} has already been estimated, say via beam scanning in radar searching mode~\cite{richards2010principles}.

With the monostatic architecture, the transmitted signal $\bar{\mathbf{X}}[n]$ is known to the ISAC receiver.
Therefore, to sense the target in the delay-Doppler domain from the received signal $\bar{\mathbf{y}}_s[n]\in\mathbb{C}^{N\times1}$ in \eqref{dam_echo}, matched filters can be constructed for each delay-Doppler bin parameterized by $(\tau_p,\nu_q)$, given by
\begin{equation}
\begin{aligned}
&\mathbf{h}^H(\tau_p,\nu_q) = \frac{\mathbf{a}^H(\theta)\mathbf{F}\bar{\mathbf{S}}[n-\tau_p]\mathrm{diag}\left(\mathbf{d}^T[\nu_q]\right)}{\left\|\mathbf{a}^H(\theta)\mathbf{F}\bar{\mathbf{S}}[n-\tau_p]\mathrm{diag}\left(\mathbf{d}^T[\nu_q]\right)\right\|},
\\
&p = 0,\cdots,P-1, q = 0,\cdots, Q-1,
\end{aligned}
\end{equation}
where $P$ and $Q$ denote the number of bins in the delay and Doppler intervals of interest, with the delay resolution of $T_s$ and Doppler frequency resolution of $1/(NT_s)$, respectively;
$\mathbf{d}[\nu_q]\triangleq\left[e^{j2\pi \nu_q(n-(N-1))T_s},\cdots,e^{j2\pi \nu_qnT_s}\right]^T\in\mathbb{C}^{N\times1}$ is defined as the Doppler frequency estimator, with $\nu_q$ denoting the Doppler frequency for the $q$th Doppler bin.
After matched filtering (MF), the resulting output is
\begin{equation}\label{mf}
\begin{aligned}
{r}(\tau_p,\nu_q) &= \mathbf{y}_s^H[n]\mathbf{h}(\tau_p,\nu_q) \\
&=\alpha\mathbf{a}^H(\theta)\mathbf{F}\bar{\mathbf{S}}[n-\tau]\mathrm{diag}\left(\mathbf{d}^T[\nu]\right)\\
&\quad\times\frac{\mathrm{diag}^H\left(\mathbf{d}^T[\nu_q]\right)\bar{\mathbf{S}}^H[n-\tau_p]\mathbf{F}^H\mathbf{a}(\theta)}{\left\|\mathbf{a}^H(\theta)\mathbf{F}\bar{\mathbf{S}}[n-\tau_p]\mathrm{diag}\left(\mathbf{d}^T[\nu_q]\right)\right\|}+\hat{z},
\end{aligned}
\end{equation}
where $\hat{z}\triangleq\frac{\mathbf{z}^H\mathrm{diag}^H\left(\mathbf{d}^T[\nu_q]\right)\bar{\mathbf{S}}^H[n-\tau_p]\mathbf{F}^H\mathbf{a}(\theta)}{\left\|\mathbf{a}^H(\theta)\mathbf{F}\bar{\mathbf{S}}[n-\tau_p]\mathrm{diag}\left(\mathbf{d}^T[\nu_q]\right)\right\|}$ is the resulting noise after MF, satisfying $\hat{z}\sim\mathcal{CN}(0,\sigma^2)$.
Therefore, the delay $\tau$ and Doppler frequency $\nu$ in \eqref{dam_echo} can be estimated as $(\hat{\tau},\hat{\nu})=\arg\max\limits_{\tau_p,\nu_q}\ \left|r(\tau_p,\nu_q)\right|$.

\subsection{Sensing performance analysis}\label{sensPerformance}
Note that the MF output in \eqref{mf} is random due to the random communication symbols involved.
To analyze the DAM sensing performance, let us define a {\it delay-Doppler correlation matrix} (DDCM) involved in \eqref{mf} as
\begin{equation}\label{corr_ele0}
\bm{\Lambda}(\tau_p,\nu_q;\tau,\nu)\triangleq\frac{1}{N}\bar{\mathbf{S}}[n-\tau]\mathrm{diag}\left(\mathbf{d}^T[\nu-\nu_q]\right)\bar{\mathbf{S}}^H[n-\tau_p].
\end{equation}
Then \eqref{mf} can be rewritten as
\begin{equation}\label{mf2}
\small
\begin{aligned}
&r(\tau_p,\nu_q;\tau,\nu) =\frac{\alpha N\mathbf{a}^H(\theta)\mathbf{F}\bm{\Lambda}(\tau_p,\nu_q;\tau, \nu)\mathbf{F}^H\mathbf{a}(\theta)}{\sqrt{N\mathbf{a}^H(\theta)\mathbf{F}\bm{\Lambda}(\tau_p,\nu_q;\tau_p,\nu_q)\mathbf{F}^H\mathbf{a}(\theta)}}+\hat{z}\\
&=\alpha\chi(\tau_p,\nu_q;\tau,\nu)\sqrt{N\mathbf{a}^H(\theta)\mathbf{F}\bm{\Lambda}(\tau_p,\nu_q;\tau_p,\nu_q)\mathbf{F}^H\mathbf{a}(\theta)}+\hat{z},
\end{aligned}
\end{equation}
where
\begin{equation}\label{AF} \chi(\tau_p,\nu_q;\tau,\nu)\triangleq\frac{\mathbf{a}^H(\theta)\mathbf{F}\bm{\Lambda}(\tau_p,\nu_q;\tau,\nu)\mathbf{F}^H\mathbf{a}(\theta)}{{\mathbf{a}^H(\theta)\mathbf{F}\bm{\Lambda}(\tau_p,\nu_q;\tau_p,\nu_q)\mathbf{F}^H\mathbf{a}(\theta)}}
\end{equation}
is termed as the discrete equivalent normalized AF of the DAM signal in delay-Doppler domain, w.r.t. the ground-truth delay $\tau$ and Doppler frequency $\nu$.
Note that according to \eqref{dam_tx3}, the element of the DDCM in \eqref{corr_ele0} at the $i$th row and $j$th column is given by
\begin{equation}\label{corr_ele}
\small
\begin{aligned}
&\left[\bm{\Lambda}(\tau_p,\nu_q;\tau,\nu) \right]_{i,j}, 1\le i,j\le L \\
&=\frac{1}{N}\mathbf{s}^T[n-\kappa_i-\tau]\mathrm{diag}\left(\mathbf{d}^T[\nu-\nu_q]\right)\mathbf{s}^\dagger[n-\kappa_j-\tau_p]\\
&=\frac{1}{N}\sum\nolimits_{\tilde{n}=n-(N-1)}^{n}s[\tilde{n}-\kappa_i-\tau]s^\dagger[\tilde{n}-\kappa_j-\tau_p]e^{j2\pi(\nu-\nu_q)\tilde{n}T_s},
\end{aligned}
\end{equation}
where $(\cdot)^\dagger$ denotes the conjugation operation.
Note that \eqref{corr_ele} involves the summation of $N$ i.i.d. communication symbols $s[\tilde{n}]$ with various delays.
When $N$ is large, we have the following result:

{\em Theorem 1}: For $N\gg 1$, the DDCM element in \eqref{corr_ele} approaches to
\begin{equation}\label{corr_ele2}
\small
\left[\bm{\Lambda}(\tau_p,\nu_q;\tau,\nu) \right]_{i,j}\rightarrow\left[\bm{\Lambda}(d_\tau,d_\nu) \right]_{i,j}=\left\{
\begin{aligned}
&\psi(d_\nu), && \kappa_i - \kappa_j = d_\tau, \\
&0, &&\kappa_i - \kappa_j \neq d_\tau,
\end{aligned}
\right.
\end{equation}
where $\bm{\Lambda}(d_\tau,d_\nu)$ only depends on the delay difference $d_\tau\triangleq \tau_p-\tau$ and the Doppler difference $d_\nu\triangleq \nu - \nu_q$.
Furthermore, $\psi(d_\nu)$ in \eqref{corr_ele2} is given by
\begin{equation}\label{Doppler}
\small
\begin{aligned}
\psi(d_\nu)&\triangleq\frac{1}{N}\sum\limits_{\tilde{n}=n-(N-1)}^n e^{j2\pi d_\nu\tilde{n}T_s}=e^{j\pi d_\nu(N-1)T_s}\frac{\sin(\pi d_\nu NT_s)}{N\sin(\pi d_\nu T_s)}\\
&=e^{j\pi d_\nu (N-1)T_s}\mathrm{asinc}(d_\nu ,N),
\end{aligned}
\end{equation}
where $\mathrm{asinc}(d_\nu ,N)\triangleq\frac{\sin(\pi d_\nu  NT_s)}{N\sin(\pi d_\nu  T_s)}$ denotes the ``aliased sinc'' function \cite{richards2010principles}.
\begin{IEEEproof}
Please refer to  Appendix \ref{appendix a}.
\end{IEEEproof}
Note that $\lim\limits_{d_\nu \rightarrow 0}\mathrm{asinc}(d_\nu ,N)=1$ and $\lim\limits_{N\rightarrow \infty}\mathrm{asinc}(d_\nu ,N)=\delta(d_\nu )$, thus we have $\psi(0) = 1$.
Therefore, according to {\it Theorem 1}, when $N\gg 1$, \eqref{mf2} approaches to
\begin{equation}\label{mf_ap}
r(d_\tau,d_\nu ) = \alpha \chi(d_\tau,d_\nu )\sqrt{N\mathbf{a}^H(\theta)\mathbf{F}\bm{\Lambda}(0,0)\mathbf{F}^H\mathbf{a}(\theta)} + \hat{z},
\end{equation}
with the AF approaching to
\begin{equation}\label{AF_ap}
\chi(d_\tau,d_\nu )  = \frac{\mathbf{a}^H(\theta)\mathbf{F}\bm{\Lambda}(d_\tau,d_\nu )\mathbf{F}^H\mathbf{a}(\theta)}{{\mathbf{a}^H(\theta)\mathbf{F}\bm{\Lambda}(0,0)\mathbf{F}^H\mathbf{a}(\theta)}},
\end{equation}
which is referred to as the asymptotic AF that is only related to the delay and Doppler differences, i.e., $d_\tau$ and $d_\nu $.
Depending on whether $d_\tau=0$ and/or $d_\nu =0$ or not, we have the following four cases.

1) $\chi(0,0)$:

In this case, both the considered delay bin $\tau_p$ and Doppler bin $\nu_q$ perfectly match with the ground-truth $\tau$ and $\nu$, respectively, i.e., $d_\tau=0$ and $d_\nu =0$.
Thus \eqref{corr_ele2} reduces to
\begin{equation}\label{pmatched1}
\begin{aligned}
&\left[\bm{\Lambda}(0,0) \right]_{i,j}=\left\{
\begin{aligned}
&1,&& \kappa_i = \kappa_j \\
&0,&& \kappa_i\neq\kappa_j
\end{aligned}
\right.
,1\le i,j\le L.
\end{aligned}
\end{equation}
Note that since by design that $\kappa_i\neq\kappa_j, \forall i\neq j$ in \eqref{dam_tx}, we have
$\left[\bm{\Lambda}(0,0) \right]_{i,j}=0,\forall i\neq j$ and $\left[\bm{\Lambda}(0,0) \right]_{i,j}=1, \forall i=j$, which indicates that the DDCM is simply an identity matrix, i.e.,
\begin{equation}\label{pmatched2}
\bm{\Lambda}(0,0)=\mathbf{I}_L.
\end{equation}
By substituting \eqref{pmatched2} into \eqref{AF_ap} and \eqref{mf_ap}, we have $\chi(0,0)=1$, and the resulting MF output is
\begin{equation}
r(0,0) = \alpha{\sqrt{N\mathbf{a}^H(\theta)\mathbf{F}\mathbf{F}^H\mathbf{a}(\theta)}}+\hat{{z}}.
\end{equation}
As a result, the maximum output SNR for DAM sensing is
\begin{equation}\label{sens_snr}
\small
\gamma= \frac{|\alpha|^2 {{N\mathbf{a}^H(\theta)\mathbf{F}\mathbf{F}^H\mathbf{a}(\theta)}}}{\sigma^2}=\frac{|\alpha|^2 {{N\mathbf{a}^H(\theta)\left(\sum_{l=1}^L\mathbf{f}_l\mathbf{f}_l^H\right)\mathbf{a}(\theta)}}}{\sigma^2}.
\end{equation}

2) $\chi(0,d_\nu )$:

In this case, the considered delay bin $\tau_p$ matches with the ground-truth $\tau$, while the Doppler bin $\nu_q$ and the ground-truth $\nu$ are mismatched,  i.e., $d_\tau=0$ and $d_\nu \neq 0$.
Thus \eqref{corr_ele2} reduces to
\begin{equation}
\begin{aligned}
&\left[\bm{\Lambda}(0,d_\nu ) \right]_{i,j}=\left\{
\begin{aligned}
&\psi(d_\nu ),&& \kappa_i = \kappa_j \\
&0,&& \kappa_i\neq\kappa_j
\end{aligned}
\right.
,1\le i,j\le L.
\end{aligned}
\end{equation}
Similar to \eqref{pmatched1} and \eqref{pmatched2}, the DDCM is a scaled identity matrix, i.e., $\bm{\Lambda}(0,d_\nu )=\psi(d_\nu )\mathbf{I}_L$, and the delay-cut of $\chi(\tau_p,\nu_q;\tau,\nu)$ at $\tau_p=\tau$ (i.e., $d_\tau=0$) approaches to
\begin{equation}\label{AF2}
\chi(0,d_\nu ) =\psi(d_\nu )=e^{j\pi d_\nu (N-1)T_s}\mathrm{asinc}(d_\nu ,N),
\end{equation}
which measures the Doppler frequency sensing performance of the DAM signal.

\begin{figure} 
  \centering
  \subfigure[Doppler-cut AF at $\nu_q=\nu=0$, i.e., $d_\nu =0$]{
  \includegraphics[width=0.48\textwidth]{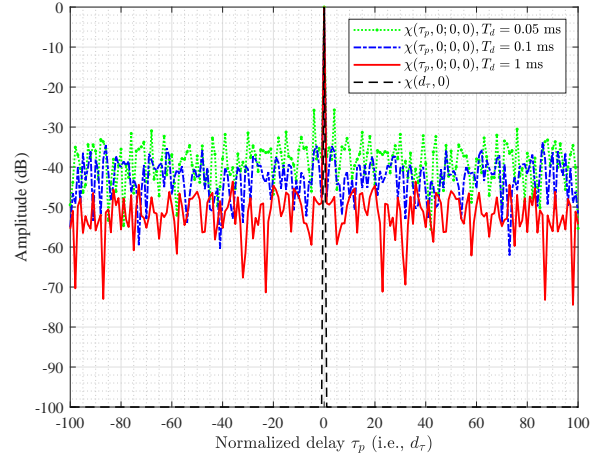}\label{DAMdoppler_cut}
  }
  \\
  \subfigure[Delay-cut AF at $\tau_p=\tau=0$, i.e., $d_\tau=0$]{
  \includegraphics[width=0.48\textwidth]{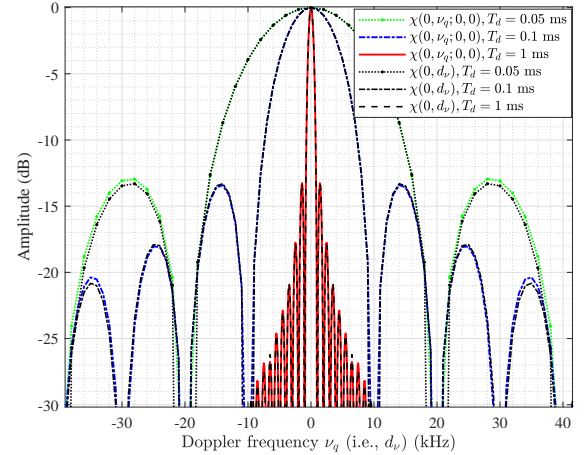}\label{DAMdelay_cut}
  }
  \caption{Comparison between the normalized AF $\chi(\tau_p,\nu_q;\tau,\nu)$ and the asymptotic AF $\chi(d_\tau,d_\nu )$ of the DAM signal with the ground-truth delay $\tau=0$ and the Doppler frequency $\nu=0$. Thus $d_\tau=\tau_p$ and $d_\nu =\nu_p$. Single-path beamforming is considered, i.e., $\mathbf{f}_1=\sqrt{\frac{P}{M}}\mathbf{a}(\theta)$ and $\mathbf{f}_l=\mathbf{0}_{M\times1}$, $\forall l\neq 1$.}\label{compr}\vspace{-10pt}
\end{figure}

For radar sensing, one important performance metric is the sidelobe level of the AF, which is critically related to the waveform sensibility for multiple targets. Specifically, the high sidelobes of the stronger target echo will mask the MF output of the weaker one \cite{richards2010principles}.
To measure the sidelobe level of $\chi(0,d_\nu )$, a common metric is the PSR, which is defined as~\cite{blunt2016overview}
\begin{equation}\label{psr_d}
\Phi_{PSR} = \max\limits_{d_\nu \neq 0}\left|{\chi(0,d_\nu )}\right|= \max\limits_{d_\nu \neq 0}\left|\mathrm{asinc}(d_\nu ,N)\right|.
\end{equation}
Note that for the asinc function, the peak sidelobe is approximately $13.2$ dB below the central peak amplitude, as illustrated in Fig.~\ref{DAMdelay_cut}, where $N$ is only related to the frequency resolution but has no effect on the sidelobe level \cite{richards2010principles}, thus we have $\Phi_{PSR} \approx -13.2$ dB.
In practice, to achieve lower PSR, some window functions can be applied, like Hamming or Taylor windows \cite{richards2010principles}.

3) $\chi(d_\tau,0)$:

In this case, the considered delay bin $\tau_p$ mismatches with the ground-truth $\tau$, while the considered Doppler bin $\nu_q$ and the ground-truth $\nu$ are matched, i.e., $d_\tau\neq 0$ and $d_\nu =0$.
Thus \eqref{corr_ele2} can be rewritten as
\begin{equation}\label{mismatched}
\begin{aligned}
&\left[\bm{\Lambda}(d_\tau,0) \right]_{i,j} =\left\{
\begin{aligned}
&1,&& \triangle_{i,j}=d_\tau, \\
&0,&& \text{otherwise},
\end{aligned}
\right.
\end{aligned}
\end{equation}
where $\triangle_{i,j}\triangleq \kappa_i - \kappa_j, 1\le i,j\le L$, is defined as the delay difference between the delay pre-compensation of the $i$th and $j$th paths.
Note that with $\kappa_l\neq\kappa_{l'}$, $\forall l\neq l'$ and $\kappa_l=n_{\max}-n_l$, we can obtain that the delay difference matrix $\mathbf{\Delta}$ consisting of $\triangle_{i,j}$, $1\le i,j\le L$, is a skew-symmetric matrix, where $\triangle_{i,j}\in~\left\{\pm1,\cdots,\pm n_d\right\}$, $ \forall i,j$, $\triangle_{i,j_1}\neq\triangle_{i,j_2}$, $\forall j_1\neq j_2$, and $\triangle_{i_1,j}\neq\triangle_{i_2,j}$, $\forall i_1\neq i_2$.
For example, for $L=3$ multi-paths with delays $n_1=1, n_2=3$, and $n_3=5$, we have $\kappa_1=4$, $\kappa_2=2$, and $\kappa_3=0$, and the delay difference matrix is
\begin{equation}\label{eg}
\mathbf{\Delta} = \left[
\begin{array}{lll}
0 & 2 & 4 \\
-2& 0 & 2 \\
-4& -2& 0 \\
\end{array}
\right].
\end{equation}
Therefore, according to \eqref{mismatched}, when $|d_\tau|>n_d$, we have $\left[\bm{\Lambda}(d_\tau,0) \right]_{i,j} = 0$, $\forall i,j$, i.e., $\bm{\Lambda}(d_\tau,0)$ is a zero matrix, which means that DAM can ensure perfect delay resolution at least for targets separated by time difference greater than the delay spread of the communication channel.
On the other hand, for $0<|d_\tau|\le n_d$, i.e., $d_\tau = \pm1,\cdots,\pm n_d$, we can obtain that $\left[\bm{\Lambda}(d_\tau,0) \right]_{i,j}=1$ if and only if $d_\tau = \triangle_{i,j}$.

To derive $\bm{\Lambda}(d_\tau,0)$ for $0<\left|d_\tau\right|\le n_d$, we first put all pairs of $(i,j)$ whose corresponding delay difference $\triangle_{i,j}$ satisfying $d_\tau = \triangle_{i,j}$ into a set
\begin{equation}\label{group}
\mathcal{S}(d_\tau)= \left\{(i,j)\mid \exists i,j, \text{s.t.}, d_\tau = \triangle_{i,j}, 1\le i,j\le L \right\}.
\end{equation}
Taking \eqref{eg} as an example, we have $\triangle_{1,2}=\triangle_{2,3}=2$. Based on \eqref{mismatched}, for $d_\tau = 2$, $\left[\bm{\Lambda}(2,0)\right]_{1,2}=\left[\bm{\Lambda}(2,0)\right]_{2,3}=1$, while all other elements of $\bm{\Lambda}(2,0)$ equal to zero.
Thus, we put the pairs of $(1,2)$ and $(2,3)$ into a set $\mathcal{S}(2)=\left\{(1,2),(2,3)\right\}$, and the DDCM when $d_\tau = 2$ for $L=3$ is given by
\begin{equation}
\bm{\Lambda}(2,0) = \left[
\begin{array}{lll}
0 & 1 & 0 \\
0 & 0 & 1 \\
0 & 0 & 0
\end{array}
\right].
\end{equation}

Therefore, according to \eqref{group}, when $0<|d_\tau|<n_d$, \eqref{mismatched} can be further written as
\begin{equation}\label{element3}
\left[\bm{\Lambda}(d_\tau,0)\right]_{i,j}= \left\{
\begin{aligned}
&1, && \text{if}\ (i,j)\in\mathcal{S}(d_\tau),\\
&0, && \mathrm{otherwise}.
\end{aligned}
\right.
\end{equation}
Furthermore, based on \eqref{group} and \eqref{element3}, it can be obtained that the Doppler-cut of $\chi(\tau_p,\nu_q;\tau,\nu)$ at $\nu_q = \nu$ (i.e., $d_\nu=0$) approaches to
\begin{equation}\label{AF_mis}
\begin{aligned}
\chi(d_\tau,0) &= \frac{\mathbf{a}^H(\theta)\mathbf{F}\bm{\Lambda}(d_\tau,0)\mathbf{F}^H\mathbf{a}(\theta)}{{\mathbf{a}^H(\theta)\mathbf{F}\bm{\Lambda}(0,0)\mathbf{F}^H\mathbf{a}(\theta)}}
\\
&=\frac{\mathbf{a}^H(\theta)\left(\sum\nolimits_{(i,j)\in\mathcal{S}(d_\tau)}\mathbf{f}_i\mathbf{f}_j^H\right)\mathbf{a}(\theta)}{{\mathbf{a}^H(\theta)\sum\nolimits_{l=1}^L\mathbf{f}_l\mathbf{f}_l^H\mathbf{a}(\theta)}},
\end{aligned}
\end{equation}
which measures the delay sensing performance of the DAM signal.
To measure the sidelobe level of $\chi(d_\tau,0)$, in addition to PSR, another effective metric is the ISR, which is particulary useful for measuring the susceptibility to distributed scattering such as clutters \cite{blunt2016overview}.
The discrete form ISR of $\chi(d_\tau,0)$ can be expressed as
\begin{equation}\label{ISR}
\begin{aligned}
&\Phi_{ISR} = {\sum\limits_{0<\left|d_\tau\right|\le n_d}\left|\chi(d_\tau,0)\right|^2}
\\
&=\frac{\sum\limits_{0<\left|d_\tau\right|\le n_d}\left|\mathbf{a}^H(\theta)\left(\sum\nolimits_{(i,j)\in\mathcal{S}(d_\tau)}\mathbf{f}_i\mathbf{f}_j^H\right)\mathbf{a}(\theta)\right|^2}{\left|{\mathbf{a}^H(\theta)\sum\nolimits_{l=1}^L\mathbf{f}_l\mathbf{f}_l^H\mathbf{a}(\theta)}\right|^2}.
\end{aligned}
\end{equation}
In general, the ISR should be no greater than a predetermined threshold to ensure the sensing performance, which can be achieved by applying the appropriate DAM path-based beamforming $\left\{\mathbf{f}_l\right\}_{l=1}^L$ as discussed in Section~\ref{DAMbeamforming}.

4) $\chi(d_\tau,d_\nu )$:

In this case, neither the delay nor the Doppler bins $(\tau_p,\nu_q)$ under consideration match with the ground-truth $(\tau,\nu)$, i.e., $d_\tau\neq 0$ and $d_\nu \neq 0$.
Therefore, according to \eqref{corr_ele2}, similar from \eqref{mismatched} to \eqref{AF_mis}, the asymptotic AF is
\begin{equation}
\begin{aligned}
\chi(d_\tau,d_\nu ) &= \frac{\psi(d_\nu )\mathbf{a}^H(\theta)\left(\sum\nolimits_{(i,j)\in\mathcal{S}(d_\tau)}\mathbf{f}_i\mathbf{f}_j^H\right)\mathbf{a}(\theta)}{{\mathbf{a}^H(\theta)\sum\nolimits_{l=1}^L\mathbf{f}_l\mathbf{f}_l^H\mathbf{a}(\theta)}}
\\
&=\chi(d_\tau,0)\chi(0,d_\nu ),
\end{aligned}
\end{equation}
which means that $\chi(d_\tau,d_\nu )$ in delay-Doppler domain can be decoupled as the product of $\chi(d_\tau,0)$ and $\chi(0,d_\nu )$ in delay and Doppler domains, respectively.

In summary, the asymptotic AF of the DAM signal with $N\gg 1$ can be written as
\begin{equation}\label{AF_final}
\chi(d_\tau,d_\nu ) = \left\{
\begin{aligned}
&1, && d_\tau = 0, d_\nu  =0\\
&\chi(d_\tau,0), && d_\tau\neq 0, d_\nu  = 0\\
&\chi(0,d_\nu ), && d_\tau =0, d_\nu \neq 0\\
&\chi(d_\tau,0)\chi(0,d_\nu ), && d_\tau\neq 0, d_\nu  \neq 0.
\end{aligned}
\right.
\end{equation}
Note that different from the delay cut AF $\chi(0,d_\nu )$ in \eqref{AF2}, the Doppler-cut AF $\chi(d_\tau,0)$ in \eqref{AF_mis} is critically dependent on the DAM beamforming vectors $\left\{\mathbf{f}_l\right\}_{l=1}^L$.
Thus, the Doppler-cut AF can be improved via appropriate beamforming for better sensing performance.
In particular, with the transmit power constraint in \eqref{power_const}, to maximize the sensing SNR in \eqref{sens_snr} while minimize the ISR of $\chi(d_\tau,0)$ in \eqref{ISR} without considering the communication performance, it only needs to retain one beam aligned with the sensing target steering vector $\mathbf{a}(\theta)$ and let other beams to zero, i.e., $\mathbf{f}_{1}=\sqrt{\frac{P_t}{M}}\mathbf{a}(\theta)$ and $\mathbf{f}_{l}=\mathbf{0}_{M\times1},\forall l\neq 1$, which is termed as {\it single-path beamforming}.
The resulting maximum sensing SNR is $\gamma_{\max}=\frac{\left|\alpha\right|^2NMP_t}{\sigma^2}$, while $\chi(d_\tau,0)=0$ and $\Phi_{ISR}=0$.
Note that for such a single-path beamforming scheme, \eqref{AF_final} reduces to
\begin{equation}
\chi(d_\tau,d_\nu ) = \left\{
\begin{aligned}
&\chi(0,d_\nu ), && d_\tau =0\\
&0, && d_\tau \neq 0
\end{aligned}
\right.
\end{equation}
which is equivalent to the asymptotic AF of the conventional SC waveform for sensing.

In Fig.~\ref{compr}, we plot the normalized AF $\chi(\tau_p,\nu_q;\tau,\nu)$ and the asymptotic AF $\chi(d_\tau,d_\nu )$ for DAM signal with the single-path beamforming.
The ground-truth delay and Doppler frequency are set as $\tau=0$ and $\nu=0$.
Thus we have $d_\tau=\tau_p$ and $d_\nu =\nu_q$.
The signals $s[n]$ are set as the $64$ quadrature amplitude modulation (QAM) symbols and the system bandwidth is $B=100$ MHz.
The number of the DAM symbols within one CPI are $N=5000, 10000, 100000$, which correspond to the CPI duration of $T_d=NT_s = 0.05, 0.1, 1$ millisecond (ms).
It is revealed that $\chi(\tau_p,\nu_q;\tau,\nu)$ approaches to $\chi(d_\tau,d_\nu )$ when $N$ is large, which verifies our theoretical result in Theorem~1.
Specifically, as shown in Fig.~\ref{DAMdoppler_cut}, for the Doppler-cut AF of $\chi(\tau_p,\nu_p;\tau,\nu)$ at $\nu_p=\nu=0$, i.e., $\chi(\tau_p,0;0,0)$, the PSR decreases as the CPI $T_d$ increases,
and $\chi(\tau_p,0;0,0)$ approaches to $\chi(d_\tau,0)$ when $T_d$ is large.
On the other hand, the delay-cut AF of $\chi(\tau_p,\nu_p;\tau,\nu)$ at $\tau_p=\tau=0$, i.e., $\chi(0,\nu_p;0,0)$, approaches to $\chi(0,d_\nu )$ for $T_d=0.05,0.1$ and $1$~ms.
Note that the PSR of $\chi(0,\nu_p;0,0)$ or $\chi(0,d_\nu )$ is always about $-13.2$ dB, while increasing the CPI $T_d$ can only improve the Doppler frequency resolution but not the PSR.

\subsection{Beamforming optimization for DAM-ISAC}\label{DAMbeamforming}
In this subsection, the beamforming optimization problem for DAM-ISAC is formulated to maximize the communication SNR in \eqref{comm_snr}, subject to the ISI zero-forcing (ZF) condition $\mathbf{h}_{l'}^H\mathbf{f}_{l}=0,\forall l\neq l'$ and the transmit power constraint in \eqref{power_const}, while guaranteeing that the sensing output SNR in \eqref{sens_snr} is no smaller than a threshold $\gamma_{th}$ and the ISR in \eqref{ISR} is no greater than a threshold $\phi_{th}$.
By discarding the constant term $\sigma^2$ in \eqref{comm_snr}, the problem can be formulated as
\begin{align}
\mathrm{(P1):}
& \max\limits_{\left\{\mathbf{f}_l\right\}_{l=1}^L} \left|\sum\nolimits_{l=1}^L\mathbf{h}_l^H\mathbf{f}_l\right|^2 \label{P1}\\
& \text{s.t.}  \quad \mathbf{h}_{l'}^H\mathbf{f}_{l}=0, \forall l\neq l', 1\le l,l'\le L,  \tag{\ref{P1}{a}}\\
&     \qquad  \sum\nolimits_{l=1}^L\left\|\mathbf{f}_l\right\|^2\le P_t, \tag{\ref{P1}{b}}\label{power_constraint}\\
& \qquad  \frac{|\alpha|^2N}{\sigma^2} {{\mathbf{a}^H(\theta)\left(\sum\nolimits_{l=1}^L\mathbf{f}_l\mathbf{f}_l^H\right)\mathbf{a}(\theta)}}\ge\gamma_{th}, \tag{\ref{P1}{c}}\label{sens_constraint}\\
&       \quad  \frac{\sum\limits_{0<\left|d_\tau\right|\le n_d}\left|\mathbf{a}^H(\theta)\bigg(\sum\limits_{(i,j)\in\mathcal{S}(d_\tau)}\mathbf{f}_i\mathbf{f}_j^H\bigg)\mathbf{a}(\theta)\right|^2}{\left|{\mathbf{a}^H(\theta)\sum\nolimits_{l=1}^L\mathbf{f}_l\mathbf{f}_l^H\mathbf{a}(\theta)}\right|^2} \le \phi_{th}. \tag{\ref{P1}{d}}\label{ISR_constraint}
\end{align}

For notational convenience, \eqref{sens_constraint} can be rewritten as
\begin{equation}\label{sens_cons}
{{\mathbf{a}^H(\theta)\left(\sum\nolimits_{l=1}^L\mathbf{f}_l\mathbf{f}_l^H\right)\mathbf{a}(\theta)}}\ge\tilde{\gamma}_{th},
\end{equation}
with $\tilde{\gamma}_{th}\triangleq  \frac{\gamma_{th}\sigma^2}{\left|\alpha\right|^2N}$.
Furthermore, to simplify the problem, we consider a more stringent ISR constraint than \eqref{ISR_constraint}, given by
\begin{equation}\label{ISR_cons}
{\sum\limits_{0<\left|d_\tau\right|\le n_d}\left|\mathbf{a}^H(\theta)\left(\sum\nolimits_{(i,j)\in\mathcal{S}(d_\tau)}\mathbf{f}_i\mathbf{f}_j^H\right)\mathbf{a}(\theta)\right|^2}\le \tilde{\phi}_{th},
\end{equation}
where $\tilde{\phi}_{th}\triangleq \phi_{th}\tilde{\gamma}_{th}^2$.
Note that due to the constraint \eqref{sens_cons}, as long as \eqref{ISR_cons} holds, \eqref{ISR_constraint} must also hold.
Therefore, problem $\mathrm{(P1)}$ can be recast as
\begin{align}
\mathrm{(P1.1)} &\max\limits_{\left\{\mathbf{f}_l\right\}_{l=1}^L}\ \left|\sum\nolimits_{l=1}^L\mathbf{h}_l^H\mathbf{f}_l\right|^2 \label{P1.1}\\
&\text{s.t.}\quad \mathbf{h}_{l'}^H\mathbf{f}_{l}=0, \forall l\neq l', 1\le l,l'\le L,  \tag{\ref{P1.1}{a}}\\
&\qquad \sum\nolimits_{l=1}^L\left\|\mathbf{f}_l\right\|^2\le P_t, \tag{\ref{P1.1}{b}}\\
&\qquad {{\mathbf{a}^H(\theta)\left(\sum\nolimits_{l=1}^L\mathbf{f}_l\mathbf{f}_l^H\right)\mathbf{a}(\theta)}}\ge\tilde{\gamma}_{th}\tag{\ref{P1.1}{c}},\\
&\  {\sum\limits_{0<\left|d_\tau\right|\le n_d}\left|\mathbf{a}^H(\theta)\bigg(\sum\nolimits_{(i,j)\in\mathcal{S}(d_\tau)}\mathbf{f}_i\mathbf{f}_j^H\bigg)\mathbf{a}(\theta)\right|^2}\le \tilde{\phi}_{th}.\tag{\ref{P1.1}{d}}\label{ISR_cons2}
\end{align}

{\it Proposition 1}: Denote by $\mathbf{H}=\left[\mathbf{h}_1,\cdots,\mathbf{h}_L\right]\in\mathbb{C}^{M\times L}$ and $\mathbf{A}(\theta)=\mathbf{a}(\theta)\mathbf{a}^{H}(\theta)\in\mathbb{C}^{M\times M}$.
Problem $\mathrm{(P1.1)}$ can be rewritten as
\begin{align}
\mathrm{(P1.2)} &\max\limits_{\mathbf{F},\left\{\mathbf{f}_l\right\}_{l=1}^L}\ \mathrm{vec}(\mathbf{F})^H\mathrm{vec}(\mathbf{H})\mathrm{vec}(\mathbf{H})^H\mathrm{vec}(\mathbf{F}) \label{P1.2}\\
&\text{s.t.}\quad \mathbf{h}_{l'}^H\mathbf{f}_{l}=0, \forall l\neq l', 1\le l,l'\le L,  \tag{\ref{P1.2}{a}}\label{ISI_cons}\\
&\qquad \mathbf{F}=\left[\mathbf{f}_1,\cdots,\mathbf{f}_L\right]\in\mathbb{C}^{M\times L} \tag{\ref{P1.2}{b}},\\
&\qquad \mathrm{vec}(\mathbf{F})^H\mathrm{vec}(\mathbf{F})\le P_t \tag{\ref{P1.2}{c}}\\
&\qquad \mathrm{vec}(\mathbf{F})^H\left(\mathbf{I}_L\otimes\mathbf{A}(\theta)\right)\mathrm{vec}(\mathbf{F})\ge\tilde{\gamma}_{th}\tag{\ref{P1.2}{d}},\\
&{\sum\limits_{0<\left|d_\tau\right|\le n_d}\left|\mathrm{vec}(\mathbf{F})^H\left(\bm{\Lambda}^T(d_\tau,0)\otimes\mathbf{A}(\theta)\right)\mathrm{vec}(\mathbf{F})\right|^2}\le \tilde{\phi}_{th},\tag{\ref{P1.2}{e}}
\end{align}
where the elements of $\bm{\Lambda}(d_\tau,0)\in\mathbb{R}^{L\times L}$ are given in \eqref{element3}, $\mathrm{vec}(\cdot)$ denotes the vectorization of a matrix, and $\otimes$ is the Kronecker product.
\begin{IEEEproof}
Please refer to  Appendix \ref{appendix b}.
\end{IEEEproof}
Let $\mathbf{H}_l=\left[\mathbf{h}_1,\cdots,\mathbf{h}_{l-1},\mathbf{h}_{l+1},\cdots,\mathbf{h}_L\right]\in\mathbb{C}^{M\times(L-1)}$.
Then \eqref{ISI_cons} can be equivalently expressed as $\mathbf{H}_l^H\mathbf{f}_l=\mathbf{0}_{(L-1)\times1},l=1,\cdots,L$, which means that $\mathbf{f}_l$ should lie in the nullspace of $\mathbf{H}_l^H$.
Denote by $\mathbf{Q}_L\triangleq\mathbf{I}_{M}-\mathbf{H}_l(\mathbf{H}_l^H\mathbf{H}_l)^{-1}\mathbf{H}_{l}^H$ the projection matrix into the space orthogonal to the columns of $\mathbf{H}_l$. Then we have $\mathbf{f}_l=\mathbf{Q}_l\mathbf{b}_l$, $l=1,\cdots,L$, where $\mathbf{b}_l\in\mathbf{C}^{M\times1}$ denotes the new vector to be optimized.
Therefore, denote by $\bar{\mathbf{b}}=\left[\mathbf{b}_1^H,\cdots,\mathbf{b}_L^H\right]^H\in\mathbb{C}^{ML\times 1}$ and $\bar{\mathbf{Q}}=\mathrm{diag}\left(\mathbf{Q}_1,\cdots,\mathbf{Q}_L\right)\in\mathbb{C}^{ML\times ML}$, then we have $\mathrm{vec}(\mathbf{F})=\bar{\mathbf{Q}}\bar{\mathbf{b}}$.
As a result, the problem $\mathrm{(P1.2)}$ can be transformed to
\begin{align}
\mathrm{(P2)}\ &\max\limits_{\bar{\mathbf{b}}}\ &&\bar{\mathbf{b}}^H\bar{\mathbf{H}}\bar{\mathbf{b}} \label{P2}\\
&\quad \text{s.t.}  &&\bar{\mathbf{b}}^H\bar{\mathbf{Q}}\bar{\mathbf{b}}\le P_t, \tag{\ref{P2}{a}}\label{power_const2}\\
&      && \bar{\mathbf{b}}^H\bar{\mathbf{A}}(\theta)\bar{\mathbf{b}}\ge\tilde{\gamma}_{th}, \tag{\ref{P2}{b}}\label{sens_cons2}\\
&      && {\sum\limits_{0<\left|d_\tau\right|\le n_d}\left|\bar{\mathbf{b}}^H\bar{\mathbf{A}}_Q(\theta,d_\tau)\bar{\mathbf{b}}\right|^2}\le \tilde{\phi},  \tag{\ref{P2}{c}} \label{ISI_cons2}
\end{align}
where $\bar{\mathbf{H}}=\bar{\mathbf{Q}}^H\mathrm{vec}(\mathbf{H})\mathrm{vec}(\mathbf{H})^H\bar{\mathbf{Q}}$, $\bar{\mathbf{Q}}=\bar{\mathbf{Q}}^H\bar{\mathbf{Q}}$ since $\mathbf{Q}_l=\mathbf{Q}_l^H\mathbf{Q}_l$, $l=1,\cdots,L$, $\bar{\mathbf{A}}(\theta)=\bar{\mathbf{Q}}^H\left(\mathbf{I}_L\otimes\mathbf{A}(\theta)\right)\bar{\mathbf{Q}}$, and $\bar{\mathbf{A}}_Q\left(\theta,d_\tau\right)=\bar{\mathbf{Q}}^H\left(\bm{\Lambda}^T(d_\tau,0)\otimes\mathbf{A}(\theta)\right)\bar{\mathbf{Q}}$.
Note that although the constraints of \eqref{power_const2} and \eqref{ISI_cons2} are convex, the objective function of $\mathrm{(P2)}$ and the left hand side (LHS) of constraint \eqref{sens_cons2} are nonconcave.
Therefore, the optimization problem is nonconvex.
Therefore, the problem is hard to tackle directly by the standard convex optimization technique.

To address this issue, we derive the SDR of $\mathrm{(P2)}$ as
\begin{align}
\mathrm{(P2.1)}\ &\max\limits_{\bar{\mathbf{B}}}\ &&\mathrm{Tr}\left(\bar{\mathbf{H}}\bar{\mathbf{B}}\right) \label{P2.2}\\
&\quad \text{s.t.}  &&\mathrm{Tr}\left(\bar{\mathbf{Q}}\bar{\mathbf{B}}\right)\le P_t, \tag{\ref{P2.2}{a}}\label{power_const_sdr}\\
&      && \mathrm{Tr}\left(\bar{\mathbf{A}}(\theta)\bar{\mathbf{B}}\right)\ge\tilde{\gamma}_{th}, \tag{\ref{P2.2}{b}}\label{sens_cons_sdr}\\
&      && {\sum\limits_{0<\left|d_\tau\right|\le n_d}\left|\mathrm{Tr}\left(\bar{\mathbf{A}}_Q(\theta,d_\tau)\bar{\mathbf{B}}\right)\right|^2}\le \tilde{\phi},  \tag{\ref{P2.2}{c}} \label{ISI_cons_sdr}\\
& &&\bar{\mathbf{B}}\succeq \mathbf{0}, \tag{\ref{P2.2}{d}} \label{sdr_cons}
\end{align}
where $\bar{\mathbf{B}}=\bar{\mathbf{b}}\bar{\mathbf{b}}^H\in\mathbb{C}^{ML\times ML}$ is the new variable to be optimized, and \eqref{sdr_cons} denotes that $\bar{\mathbf{B}}$ is positive semidefinite.
Note that the rank-one constraint of $\bar{\mathbf{B}}$ is omitted, and the LHS of the constraint \eqref{ISI_cons_sdr} is now a quadratic constraint.
Thus problem $\mathrm{(P2.1)}$ becomes a standard convex optimization problem, which can be efficiently solved by using the convex optimization tools, like CVX \cite{luo2010semidefinite}.
When the obtained optimal solution $\bar{\mathbf{B}}^{\star}$ of $(\mathrm{P2.1})$ has rank greater than one, we may obtain a feasible solution $\bar{\mathbf{b}}^{\star}$ with some rank-reduction techniques, like eigenvalue decomposition and Gaussian randomization \cite{luo2010semidefinite}.

\section{DAM Versus OFDM For Sensing}\label{DAMversusOFDM}
\begin{figure} 
  \centering
  \subfigure[DAM block structure]{
  \includegraphics[width=0.38\textwidth]{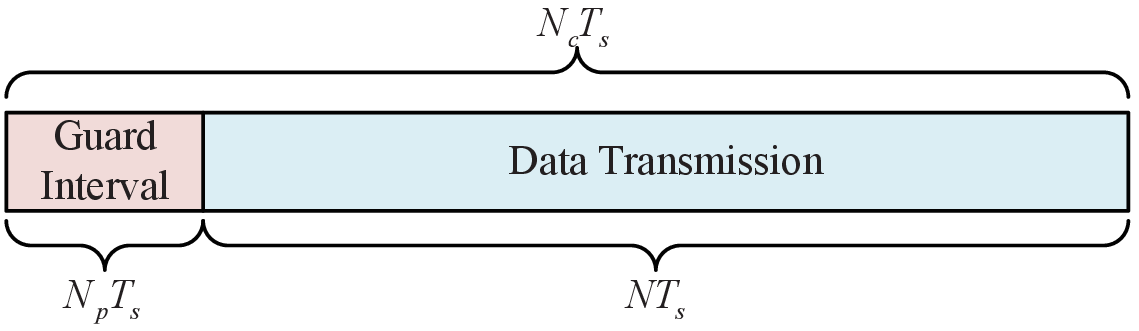}\label{damb}
  }
  \\
  \subfigure[OFDM block structure]{
  \includegraphics[width=0.38\textwidth]{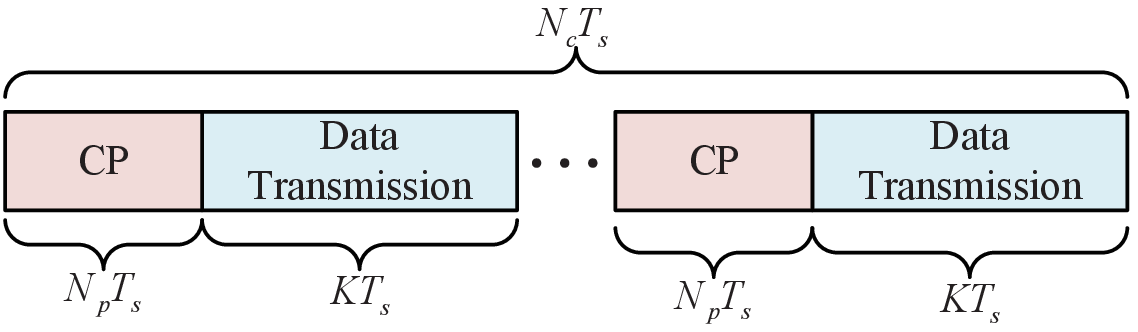}\label{ofdm}
  }
  \caption{An illustration of DAM and OFDM block structures \cite{lu2022delay}.}\label{block}\vspace{-10pt}
\end{figure}
In this section, we compare the sensing performance between DAM and OFDM, while their comparison on communication performance has been given in \cite{lu2022delay} and \cite{lu2022delay2}.
The block structures of DAM and OFDM schemes are illustrated in Fig.~\ref{block}.
Denote by $T_c=N_cT_s$ the channel coherent time, during which both the communication channel and target's states are assumed to be unchanged.
For DAM, to eliminate the inter-block interference (IBI), we use a guard interval with the time duration $T_p=N_pT_s$ for each coherent block.
Note that the length of the guard interval (i.e. $N_p$) should be set by jointly considering the communication and sensing performance requirements.
On one hand, for communication, $N_p$ should be no smaller than the maximum delay over all coherent blocks, i.e., $N_p\ge\tilde{n}_{\max}$, with $\tilde{n}_{\max}\ge n_{\max}$.
On the other hand, for sensing, the IBI may cause the range estimation ambiguity. The maximum unambiguity range is $R_{ua}=\frac{cN_pT_s}{2}$.
Thus, for DAM-ISAC, $N_p$ should be set as $N_p=\max\left\{\tilde{n}_{\max},\frac{2R_{ua}}{cT_s}\right\}$.
Therefore, as illustrated in Fig.~\ref{damb}, the block structure of DAM satisfies $N_c=N+N_p$, where $N$ denotes the number of DAM symbols for each block.

On the other hand, for OFDM with $K$ subcarriers, the subcarrier spacing is $\triangle f = 1/(KT_s)$ and the OFDM symbol duration is $T_b=1/\triangle f = KT_s$.
Denote the length of cyclic prefix (CP) as $T_p=N_pT_s$.
Thus the OFDM block structure is shown in Fig.~\ref{ofdm}, with $N_c=(N_p+K)I$, where $I>1$ denotes the number of OFDM symbols per block.
It can be inferred from Fig.~\ref{block} that $N=KI+N_p(I-1)>KI$, thanks to the saving of guard interval with DAM transmission.

In the following, we compare the sensing performance of DAM and OFDM in terms of ambiguity and resolution, PAPR, and the sensing SNR performance for delay and Doppler frequency estimation.

\subsection{Ambiguity and resolution}\label{ambiguity}
For a MISO-OFDM system with $M$ transmit antennas, the discrete-time equivalent of the $i$th transmitted OFDM symbol can be written as
\begin{equation}
\begin{aligned}
\mathbf{x}_i[n] = \frac{1}{\sqrt{K}}\sum\nolimits_{k=0}^{K-1}\mathbf{w}_kX_i[k]e^{j2\pi kn/K},&\\
0\le n\le K-1, 0\le i\le I-1,&
\end{aligned}
\end{equation}
where $\mathbf{w}_k\in\mathbb{C}^{M\times 1}$ denotes the transmit beamforming vector for the $k$th subcarrier, $X_i[k]$ is the modulated information symbol at the $k$th subcarrier for the $i$th OFDM symbol, which is assumed to belong to a finite alphabet $\mathcal{A}$, with the normalized power $\mathbb{E}[|X_i[k]|^2]=1$ and the maximum amplitude $A_{\max}=\max_{X_i[k]\in\mathcal{A}}\left|X_i[k]\right|$.

\begin{figure} 
  \centering
  \subfigure[Range profiles of OFDM and DAM sensing]{
  \includegraphics[width=0.48\textwidth]{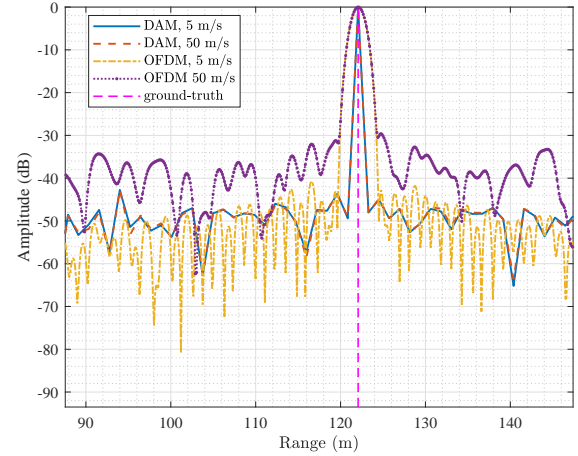}\label{doppler_cut}
  }
  \\
  \subfigure[Doppler profiles of OFDM and DAM sensing]{
  \includegraphics[width=0.48\textwidth]{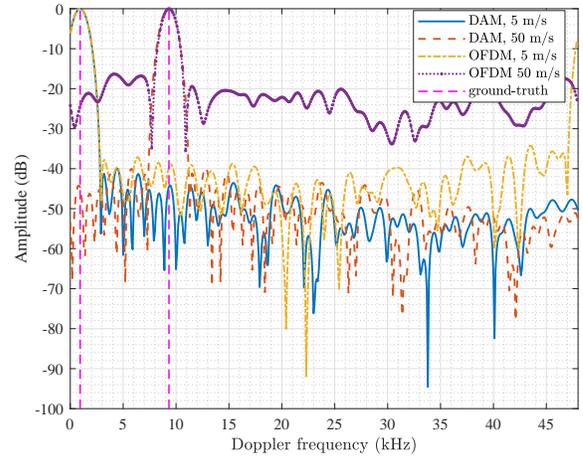}\label{delay_cut}
  }
  \caption{Comparison between DAM and OFDM on the performance of range and Doppler frequency estimation. A target at the range of $122$ m under radial velocities of $5$ m/s and $50$ m/s are considered. Both the DAM and OFDM systems operate at the carrier frequency of $28$ GHz with the bandwidth of $122.88$ MHz. Moreover, single-path beamforming is applied for DAM sensing, while all subcarrier beamforming vectors of OFDM are aligned with the steering vector of the target. Note that for OFDM sensing and DAM Doppler estimation, the Hamming window is applied to reduce the sidelobe level.}\label{DAMvsOFDM}\vspace{-10pt}
\end{figure}

For OFDM sensing, we consider the standard FFT-based signal processing in \cite{sturm2011waveform} as a benchmark.
Therefore, to sense a target with the channel given in \eqref{radar_model}, after CP removal, the $i$th received OFDM symbol can be written as
\begin{equation}
\small
\begin{aligned}
&r_i[n] = \alpha\mathbf{a}^H(\theta)\mathbf{x}_i[n-\tau]e^{j2\pi (iT_o+nT_s)\nu} + z_i[n]\\
&=\alpha\mathbf{a}^H(\theta)\frac{1}{\sqrt{K}}\sum\limits_{k=0}^{K-1}\mathbf{w}_kX_i[k]e^{j2\pi k(n-\tau)/K}e^{j2\pi (iT_o+nT_s)\nu}+z_i[n]\\
&=\frac{\alpha\mathbf{a}^H(\theta)}{\sqrt{K}}\sum\limits_{k=0}^{K-1}\mathbf{w}_kX_i[k]e^{j2\pi\left(\frac{\nu}{\triangle f}+k\right)\frac{n}{K}}e^{-j2\pi\tau \frac{k}{K}}e^{j2\pi iT_o\nu} + z_i[n],
\end{aligned}
\end{equation}
where $T_o=(N_p+K)T_s$ is the total OFDM symbol duration including the CP, and $z_i[n]$ denotes the AWGN, with $z_i[n]\sim\mathcal{CN}(0,\sigma^2)$.
With the assumption of $|\nu|\ll \triangle f$, the discrete fourier transform (DFT) of $r_i[n]$ can be written as
\begin{equation}\label{OFDM_echo}
\begin{aligned}
r_i[k] = \alpha\mathbf{a}^H(\theta)\mathbf{w}_kX_i[k]e^{j2\pi iT_o\nu}e^{-j2\pi k\tau/K} + z_i[k],&\\
0\le i\le I-1, 0\le k\le K-1,&
\end{aligned}
\end{equation}
where $z_i[k]$ denotes the DFT of $z_i[n]$, satisfying $z_i[k]\sim\mathcal{CN}(0,\sigma^2/K)$.
Then, according to \cite{sturm2011waveform}, after element-wise division, i.e., $\hat{r}_i[k]=r_i[k]/X_i[k]$,
by applying the the IFFT and FFT operations on $\hat{r}_0[k]$, $0\le k\le K-1$ and $\hat{r}_i[0], 0\le i\le I-1$, respectively, the delay and Doppler frequency of the target can be estimated as
\begin{equation}\label{OFDM_est}
\begin{aligned}
&\hat{\tau}=\arg\max\limits_{{\tau_p}}\sum\nolimits_{k=0}^{K-1}\hat{r}_0[k]e^{j2\pi k{\tau_p}/K}, \\
& \hat{\nu}=\arg\max\limits_{{\nu_q}}\sum\nolimits_{i=0}^{I-1}\hat{r}_i[0]e^{-j2\pi iT_o{\nu_q}},
\end{aligned}
\end{equation}
with the delay resolution of $T_s$ and Doppler frequency resolution of $1/T_c$, which is the same as that of DAM sensing.
Note that such a FFT-based method works well when the element-wise division operation uses the correct $X_i[k]$ from $r_i[k]$, which requires that $0\le\tau\le N_p$ and $\left|\nu\right|\ll \triangle f$.
However, the assumption of $\left|\nu\right|\ll \triangle f$ may not hold for high mobility scenarios and/or with high carrier frequency.
As a concrete example, consider a OFDM-ISAC system operated at the 5G NR FR2~\cite{TS38.101}.
With the carrier frequency of 28 GHz and the subcarrier spacing of 60 kHz. A target with radial velocity 50 m/s (180 km/h) can cause the Doppler frequency about $|\nu|=9.33$ kHz, with the $|\nu|/\triangle f\approx~15.55\%$.
Therefore, \eqref{OFDM_echo} does not hold for such scenarios, and the Doppler-dependent phase-shift across fast-time samples of each OFDM symbol may cause the high PSR and hence degrades the OFDM sensing performance.
To visualize this effect, Fig.~\ref{DAMvsOFDM} plots the delay and Doppler frequency profiles of  OFDM radar to sense a target at the range of $122$ meter (m) under two different radial velocities of $5$ m/s and $50$ m/s, respectively, where Hamming window is applied to reduce the sidelobes of the OFDM sensing outputs \cite{sturm2011waveform}.
It can be observed that although the peaks of profiles can match with the ground-truth range and Doppler frequencies, the PSRs of OFDM radar increase severely as the target velocity increases, and hence resulting in degraded sensing performance for high mobility scenarios.

On the other hand, for the proposed DAM-based sensing, to avoid the IBI across different blocks, it is also assumed that $0\le \tau\le N_p$, and the first $N_p$ elements corresponding to the guard interval are discarded.
Thus, the delay and Doppler frequency of the target can be estimated based on the remaining signals as discussed in Section~\ref{sensPerformance}.
Therefore, it can be concluded that DAM has theoretically the same maximum unambiguity range, range resolution, and velocity resolution as that of OFDM.
However, DAM as a SC waveform is more robust to the Doppler frequency estimation, where the unambiguity maximum Doppler frequency estimation is on the order of the system bandwidth $B$, rather than $B/K$ in OFDM system.
Specifically, as shown in Fig.~\ref{DAMvsOFDM}, the PSR of the DAM significantly  outperforms OFDM in high mobility scenarios.
Moreover, for OFDM sensing, due to the entailed Fourier transform, the PSRs of the resulted output are around $-13$ to $-15$ dB \cite{sturm2011waveform}.
Therefore, to mitigate this issue, Hamming window is typically applied, but at the cost of the widened main-lobe and degraded resolution.
However, for DAM-based sensing, with the MF processing, the sidelobe level of the MF output is only related to the CPI duration, which is evident from Fig.~\ref{DAMdoppler_cut}.
Therefore, the Hamming window is not needed, and DAM can achieve better range resolution than OFDM in practice, as observed from Fig.~\ref{doppler_cut}.

\subsection{PAPR and SNR performance}
In this subsection, we compare the PAPR of OFDM with that of DAM, and study the effect of transmit power backoff for sensing SNR performance.
For a multi-antenna system, the PAPR is defined as~\cite{hung2014papr}
\begin{equation}\label{PAPR}
\mathrm{PAPR} = \max\limits_{0\le m\le M-1}\left\{\mathrm{PAPR}^{(m)}\right\},
\end{equation}
where $\mathrm{PAPR}^{(m)}$ denotes the PAPR of the $m$th transmit antenna, which is given by
\begin{equation}\label{PAPR_m}
\mathrm{PAPR}^{(m)} = \max\limits_{0\le t\le T_{ob}}\frac{\left|x^{(m)}(t)\right|^2}{\mathbb{E}\left[\left|x^{(m)}(t)\right|^2\right]},
\end{equation}
where $x^{(m)}(t), 0\le t\le T_{ob}$, denotes the transmit signal at the $m$th antenna within the time interval $[0,T_{ob})$.
In general, to study the PAPR of multi-antenna systems, one needs to take into account the spectral pulse shaping and beamforming techniques, and examine the continuous-time transmitted signal.
Here, to simplify the analysis, we consider a Nyquist pulse shaping and study the PAPR for the symbol-rate sampled discrete-time signal.

For the MISO-OFDM system with $M$ transmit antennas, the discrete-time equivalent of the $i$th transmitted OFDM symbol at the $m$th antenna can be expressed as
\begin{equation}\label{OFDM_tx_m}
\begin{aligned}
x_i^{(m)}[n]=\frac{1}{\sqrt{K}}\sum\nolimits_{k=0}^{K-1}w_k^{(m)}X_i[k]e^{j2\pi kn/K},& \\
1\le n\le K-1, 0\le i\le I-1,&
\end{aligned}
\end{equation}
where $w_k^{(m)}, m=1,\cdots,M$, denotes the $m$th element of the transmit beamforming vector $\mathbf{w}_k$ at the $k$th subcarrier.
According to \eqref{PAPR} and \eqref{PAPR_m}, the PAPR of the discrete-form OFDM signal can be expressed as
\begin{equation}\label{PAPR_m1}
\small
\begin{aligned}
&\mathrm{PAPR}_{\text{OFDM}} = \max_{\substack{0\le n\le K-1,\\ 0\le i \le I-1,\\ 0\le m\le M-1}}\left\{\frac{\left|\sum\nolimits_{k=0}^{K-1}w_k^{(m)}X_i[k]e^{j2\pi kn/K}\right|^2}{\mathbb{E}\left[\left|\sum\nolimits_{k=0}^{K-1}w_k^{(m)}X_i[k]e^{j2\pi kn/K}\right|^2\right]}\right\}\\
&\overset{(a)}{=} \frac{\left(\sum\nolimits_{k=0}^{K-1}\left|w_k^{(\hat{m})}X_{\hat{i}}[k]\right|\right)^2}{\sum\nolimits_{k=0}^{K-1}\left|w_k^{(\hat{m})}\right|^2}\overset{(b)}{=}\frac{A_{\max}^2\left(\sum\nolimits_{k=0}^{K-1}\left|w_k^{(\hat{m})}\right|\right)^2}{\sum\nolimits_{k=0}^{K-1}\left|w_k^{(\hat{m})}\right|^2},
\end{aligned}
\end{equation}
where $(a)$ holds when all terms $w_k^{(\hat{m})}X_{\hat{i}}[k]e^{j2\pi k\hat{n}/K}$ are added coherently for some $(\hat{n},\hat{i},\hat{m})$, while $(b)$ holds when $\left|X_{\hat{i}}[k]\right|=A_{\max}$, $\forall k$.

\begin{figure} 
  \centering
  \includegraphics[width=0.48\textwidth]{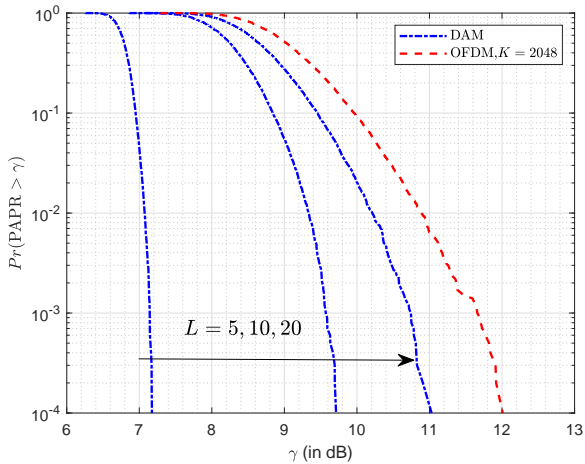}
  \caption{A comparison on the complementary cumulative distribution functions (CCDFs) of PAPR of OFDM signal with $K=2048$ subcarriers and DAM signal with $L=5,10,20$ paths for quadrature phase-shift keying (QPSK) modulation, where we assume that the transmit beamforming elements have constant modulus.}\label{figPAPR}\vspace{-0.6cm}
\end{figure}

On the other hand, for DAM, the signal transmitted by the $m$th antenna can be written as
\begin{equation}
\small
x^{(m)}[n] = \sum\nolimits_{l=1}^L f_l^{(m)}s[n-\kappa_l], 0\le n\le N-1,
\end{equation}
where $f_l^{(m)}$ denotes the $m$th element of the transmit beamforming vector $\mathbf{f}_l$.
For fairness, the same symbol constellation set is used, i.e., $s[n]\in\mathcal{A}$, $\mathbb{E}[|s[n]|^2]=1$, and $A_{\max}=\max_{s[n]\in\mathcal{A}}\left|s[n]\right|$
Therefore, similar to the OFDM case, the PAPR of DAM can be expressed as
\begin{equation}
\small
\begin{aligned}
\mathrm{PAPR}_{\text{DAM}}
&=\max_{\substack{0\le n\le N-1,\\ 0\le m\le M-1}}\left\{\frac{\left|\sum\nolimits_{l=1}^L f_l^{(m)}s[n-\kappa_l]\right|^2}{\mathbb{E}\left[\left|\sum\nolimits_{l=1}^L f_l^{(m)}s[n-\kappa_l]\right|^2\right]}\right\}\\
&\overset{(c)}{=}\frac{A_{\max}^2\left(\sum\nolimits_{l=1}^{L}\left|f_l^{(m^\star)}\right|\right)^2}{\sum\nolimits_{l=1}^{L}\left|f_l^{(m^\star)}\right|^2},
\end{aligned}
\end{equation}
where $(c)$ holds when $f_l^{(m^\star)}s[n^\star-\kappa_l]$, $1\le l\le L$ are added coherently for some $(n^\star,m^\star)$ and $\left|s[n^\star-\kappa_l]\right|=A_{\max}$, $1\le l\le L$, which is similar to \eqref{PAPR_m1}.
However, the critical difference is that for DAM, there is only $L$ signals added coherently instead of $K$ as for OFDM.
For mmWave channels with multi-path sparsity when $L\ll K$, it is expected that DAM has lower PAPR than OFDM.
Specifically, for phase-shift keying (PSK) modulation, i.e., $A_{\max}=1$, if each of transmit beamforming element has constant modulus,
say $|w_k^{(m)}|=C_1$, $\forall m,k$ and $|f_l^{(m)}|=C_2$, $\forall m,l$, where $C_1$ and $C_2$ are constants, we can obtain that the PAPRs for OFDM and DAM signals are $\mathrm{PAPR}_{\text{OFDM}} = K$ and $\mathrm{PAPR}_{\text{DAM}} = L$, respectively.
In Fig.~\ref{figPAPR}, a more informative comparison between DAM and OFDM on the distribution of the PAPR is given, from which it can be observed that DAM outperforms OFDM in terms of PAPR when the communication channel is sparse, i.e.,  $L\ll K$.

Since the nonlinear distortion may be caused when the amplifier is saturated,  power backoff is necessary if the PAPR of the transmitted signal is large.
However, this may compromise the SNR performance.
Specifically, denote by $P_{\max}$ the maximum allowable peak power for both OFDM and DAM, beyond which nonlinear distortion may be caused.
By considering the case of PSK baseband modulation and constant beamforming modulus, we have  $\mathrm{PAPR}_{\text{OFDM}}=K$ and $\mathrm{PAPR}_{\text{DAM}}=L$.
Therefore, the average transmit powers for OFDM and DAM are $P_t'=\frac{P_{\max}}{\mathrm{PAPR}_{OFDM}}=\frac{P_{\max}}{K}$ and $P_t=\frac{P_{\max}}{\mathrm{PAPR}_{DAM}}=\frac{P_{\max}}{L}$, respectively.
For OFDM-based radar sensing, it can be obtained from \eqref{OFDM_echo} and \eqref{OFDM_est} that after signal processing, for the particular delay-Doppler bin where the target lies, the output SNR is
$
\gamma_{\text{OFDM}} = \frac{|\alpha|^2I\sum\nolimits_{k=1}^K\left\|\mathbf{a}^H(\theta)\mathbf{w}_k\right\|^2}{\sigma^2/K}.
$
With the target direction $\theta$ known, the maximum output SNR of OFDM sensing is
\begin{equation}
\gamma_{\text{OFDM},\max}=\frac{|\alpha|^2MIKP_t'}{\sigma^2}=\frac{|\alpha|^2MIP_{\max}}{\sigma^2}
\end{equation}
which can be obtained by setting $\mathbf{w}_k=\sqrt{\frac{P_t'}{KM}}\mathbf{a}(\theta),\forall k$.
On the other hand, for DAM-based sensing, according to \eqref{sens_snr}, the maximum output SNR is
\begin{equation}
\gamma_{\text{DAM},\max}=\frac{|\alpha|^2MNP_t}{\sigma^2}=\frac{|\alpha|^2MNP_{\max}}{L\sigma^2}
\end{equation}
which can be obtained by setting  $\mathbf{f}_l=\sqrt{\frac{P_t}{LM}}\mathbf{a}(\theta),\forall l$.
Note that typically, we have $N>KI$ and for sparse channels, $L\ll K$.
Thus we have $N/L\gg N/K>I$ and $\gamma_{\text{DAM},\max}\gg\gamma_{\text{OFDM},\max}$, which implies that DAM can achieve much greater SNR than OFDM, thanks to the saving of guard interval and lower PAPR,

\section{Numerical results}\label{numerical}
In this section, simulation results are provided to evaluate the performance of the proposed DAM technique for ISAC as shown in Section~\ref{DAM performance}.
The carrier frequency is $f_c = 28$ GHz, and the system bandwidth is $B = 100$ MHz, which corresponds to the temporal resolution of $T_s=1/B = 10$ nanosecond (ns) and the range solution of $\triangle R=\frac{cT_s}{2}=1.5$~m.
The channel coherent time is $T_c = 1$ ms, and the guard interval is $T_p = 4$ microseconds ($\upmu$s), which corresponds to the maximum unambiguity range of $R_{\text{ua}}=600$~m.
The transmit power is $P_t = 30$ dBm, and the noise power is $\sigma^2=-89$~dBm.
The transmitter is equipped with an uniform linear array (ULA) consisting of $M=64$ antennas.
For millimeter wave (mmWave) communication, it is assumed that the channel is sparse with the number of temporal resolvable multi-paths of $L=3$.
The communication distance is $R_c=100$ m, and the channel of each delay path is modelled as $\mathbf{h}_l=\beta_l\sum\nolimits_{i=1}^{\mu_{l}}v_{li}\mathbf{a}(\theta_{li})$, where $\beta_l$ denotes the complex channel coefficient of the $l$th path, which is modelled by following \cite{zeng2016millimeter} and \cite{akdeniz2014millimeter}, while $\mu_l$ denotes the number of sub-paths for the $l$th path with same delay but different AoDs $\theta_{li},i=1,\cdots, \mu_l$; $v_{li}$ is the complex coefficient of the $i$th sub-path for the $l$th path, with $v_{li}=\frac{1}{\sqrt{\mu_l}}e^{j\phi_{li}}$ and $\phi_{li}$ satisfying the uniformly distribution in $[0,2\pi)$ \cite{lu2022delay}.
We assume that $\mu_l$ is uniformly distributed in $[1,\mu_{\max}]$, with $\mu_{\max}=3$, while the AoDs are distributed within $[-50^{\circ},50^{\circ}].$
On the other hand, for radar sensing, we consider a clear LoS path between the ISAC node and target, and the two-way propagation gain is modelled as $|\alpha|^2=\frac{\lambda^2\xi}{(4\pi)^3R^4}$, where $R=225$ m is the sensing distance and $\xi=1$ $\text{m}^2$ is the RCS of the target.

\begin{figure} 
  \centering
  \includegraphics[width=0.48\textwidth]{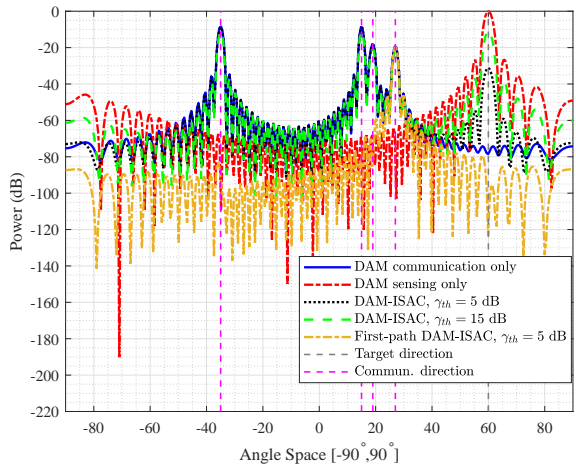}
  \caption{Normalized transmit beampatterns for DAM communication only, DAM sensing only, and DAM-ISAC, where the AoDs of communication multi-paths are $[-35^\circ,15^\circ,19^\circ,27^\circ]$, while the direction of the sensing target is $\theta=60^\circ$ with two different sensing SNR constraints of $\gamma_{th}=5$ and $15$ dB, and the ISR constraint is $\phi_{th}=-40$ dB.}\label{DAMbeampattern}\vspace{-0.35cm}
\end{figure}
Fig.~\ref{DAMbeampattern} shows the transmit beampatterns of different beamforming schemes for ``DAM communication only'', ``DAM sensing only'', and ``DAM-ISAC''.
It can be observed that for DAM communication without considering the target sensing, it only forms beams to match with the communication multi-path channels at the AoDs of $[-35^\circ,15^\circ,19^\circ,27^\circ]$, while for DAM sensing only beamforming, the formed beam only points towards the direction of the sensing target at $\theta=60^\circ$.
By contrast, for DAM-ISAC when both communication and sensing performance are considered, the proposed beamforming scheme can simultaneously generate beams pointing towards all the directions of communication and sensing channels.
Thus, it can simultaneously sense the target while providing ISI-free communication for communication user.
To further illustrate this point, the DAM-ISAC beampattern for the first path beamforming $\mathbf{f}_1$ is also shown in Fig.~\ref{DAMbeampattern} by the yellow dot line, where the resulting beam only points towards the direction of the first path, while suppressed towards other multi-path directions to ensure ISI-free communication.

\begin{figure} 
  \centering
  \includegraphics[width=0.48\textwidth]{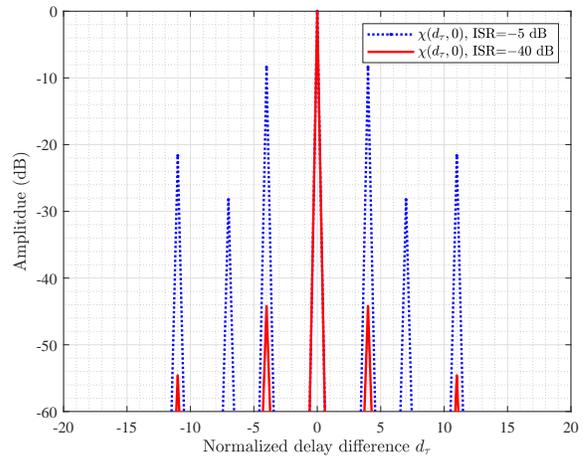}
  \caption{Normalized Doppler-cut AF for different ISR constraints $\phi_{th}$, where the $\phi_{th}=-5$ and $-40$ dB and the sensing SNR threshold is $\gamma_{th}=15 $ dB. The delays of multi-paths are $n_1=7$, $n_2=18$, $n_3=11$, thus we have $\kappa_1=11$, $\kappa_2=0$, $\kappa_3=7$, and the delay differences are $\left\{\pm4,\pm 7, \pm 11\right\}$. }\label{ISRconsAF}\vspace{-0.5cm}
\end{figure}
Fig.~\ref{ISRconsAF} gives the normalized Doppler-cut AF of the DAM signal with the proposed DAM-ISAC beamforming by solving the problem $(\mathrm{P1})$ under two ISR threshold constraints, i.e., $\phi_{th}=-5$ and $-40$~dB, while the sensing SNR threshold is set as $\gamma_{th}=15$~dB.
The delays of communication multi-paths are $n_1=7$, $n_2=18$, $n_3=11$.
Thus according to \eqref{AF_mis}, without the ISR constraint in \eqref{ISR_constraint}, the normalized Doppler-cut AF will have high sidelobes at the delay differences of $\left\{\pm4,\pm7,\pm11\right\}$, and hence rendering the degraded delays sensing performance.
Fortunately, such an issue can be effectively mitigated by introducing the ISR constraint \eqref{ISR_constraint} in the optimization problem $\mathrm{(P1)}$.
It can be observed from Fig.~\ref{ISRconsAF} that when the ISR constraint increases from $-5$ dB to $-40$ dB, the sidelobe level of the Doppler-cut AF can be significantly reduced.
Moreover, when $\phi_{th}=-40$~dB, it can be observed that the DAM-ISAC scheme can achieve comparable delay sensing performance as that of the single-path beamforming as shown in Fig.~\ref{doppler_cut}, where the sidelobe levels are below $-40$ dB.

\begin{figure} 
  \centering
  \includegraphics[width=0.48\textwidth]{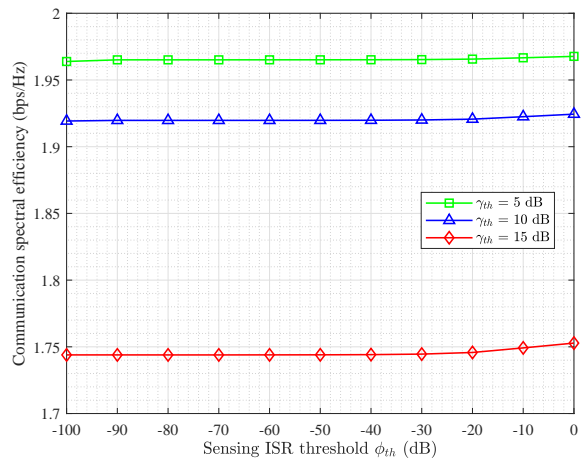}
  \caption{Average communication spectral efficiency versus sensing ISR threshold $\phi_{th}$ for DAM-ISAC beamforming with three different sensing SNR thresholds $\gamma_{th}=5, 10$, and $15$ dB.}\label{RateversusISR}\vspace{-0.8cm}
\end{figure}
To analyze the impact of the sensing ISR threshold $\phi_{th}$ on the performance of communication, Fig.~\ref{RateversusISR} shows the average communication spectral efficiency versus the sensing ISR threshold $\phi_{th}$ for three different sensing SNR thresholds, i.e., $\gamma_{th}=5, 10$, and $15$~dB, where according to the DAM block structure given in Section~\ref{DAMversusOFDM}, the communication spectral efficiency is defined as $C_{\text{DAM}} = \frac{N}{N_c}\log_2(1+\gamma_c^\star)$ \cite{lu2022delay}, with $\gamma_c^\star$ denoting the output communication SNR obtained by solving the optimization problem $\mathrm{(P1)}$.
It can be observed that changing $\phi_{th}$ only slightly affects the communication spectral efficiency.
This is appealing, since as shown in Fig.~\ref{ISRconsAF}, the ISR constraint of $-40$ dB can result in a good sensing performance with low sidelobe levels.
Therefore, the proposed DAM-ISAC beamforming scheme can achieve low sidelobe levels with negligible degradation on communication rate.
However, different from the ISR threshold $\phi_{th}$, the communication spectral efficiency obviously decreases when the sensing SNR requirements increases from $5$ to $10$ and $15$ dB.
This is evident from Fig.~\ref{DAMbeampattern}, since as the sensing SNR constraint increases from $5$ to $15$ dB, more transmit power directs towards the sensing target, rendering lower communication rate.

\begin{figure} 
  \centering
  \includegraphics[width=0.48\textwidth]{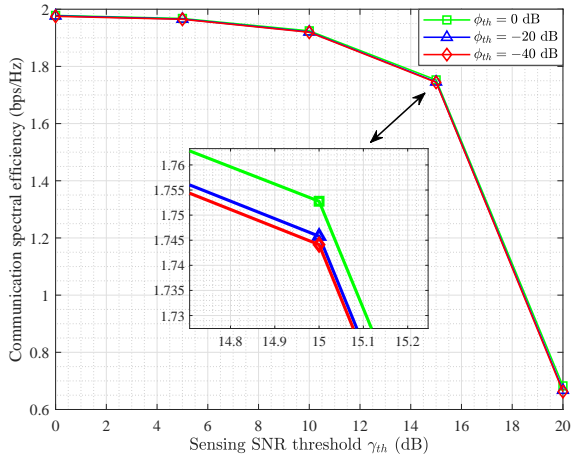}
  \caption{Average communication spectral efficiency versus sensing SNR threshold $\gamma_{th}$ for DAM-ISAC beamforming with different sensing ISRs, $\phi_{th}=0, -20$, and $-40$ dB.}\label{RateversusSNR}\vspace{-0.6cm}
\end{figure}
As a further illustration, Fig.~\ref{RateversusSNR} shows the average communication spectral efficiency versus the sensing SNR threshold $\gamma_{th}$ for different sensing ISR constraints $\phi_{th}=-40, -20$, and $0$ dB.
It can be observed that there is a clear trade-off between sensing SNR and communication spectral efficiency, i.e., the communication spectral efficiency decreases as the sensing SNR threshold $\gamma_{th}$ increases, while varying the ISR threshold $\phi_{th}$ from $0$ dB to $-40$ dB only slightly reduces the communication spectral efficiency.

\section{Conclusion}\label{conclude}
This paper studied ISAC with the novel DAM technique, which is an equalization-free SC transmission scheme that exploits the high spatial dimension and multi-path sparsity of mmWave/Terhertz massive MIMO channel.
Compared with OFDM, DAM scheme as a SC transmission scheme has lower PAPR, reduced guard interval overhead, and robust to CFO, thus DAM may achieve better communication and sensing performance.
Simulation results demonstrated that the proposed DAM-ISAC beamforming method can simultaneously provide ISI-free communication with high spectral efficiency while guaranteeing the sensing performance in terms of the sensing SNR and ISR.

\begin{appendices}
\section{Proof of Theorem 1}\label{appendix a}
Define $x[\tilde{n}] \triangleq s[\tilde{n}-\kappa_i-\tau]s^\dagger[\tilde{n}-\kappa_j-\tau_p]$ and $e^{j\varphi_{\tilde{n}}}\triangleq e^{j2\pi(\nu-\nu_q)\tilde{n}T_s}=e^{j2\pi d_\nu \tilde{n}T_s}$. Without loss of generality, by considering $n=N$, then \eqref{corr_ele} can be rewritten as
\begin{equation}\label{Y_N}
\left[\bm{\Lambda}(\tau_p,\nu_q;\tau,\nu)\right]_{i,j} =\frac{1}{N}\sum\limits_{\tilde{n}=1}^N x[\tilde{n}]e^{j\varphi_{\tilde{n}}}.
\end{equation}
Denote by $Y_N \triangleq \left[\bm{\Lambda}(\tau_p,\nu_q;\tau,\nu)\right]_{i,j}$, thus the expectation of $Y_N$ is given by
\begin{equation}\label{eY_N}
\mathbb{E}\left[Y_N\right] = \mathbb{E}\left[\frac{1}{N}\sum\nolimits_{\tilde{n}=1}^Nx[\tilde{n}]e^{j\varphi_{\tilde{n}}}\right] = \frac{1}{N}\sum\nolimits_{\tilde{n}=1}^N\mathbb{E}\left[x[{\tilde{n}}]\right]e^{j\varphi_{\tilde{n}}},
\end{equation}
It will be derived later in \eqref{eY_N1} and \eqref{eY_N2} that $\mathbb{E}\left[Y_N\right] =~\left[\bm{\Lambda}(d_\tau,d_\nu )\right]_{i,j}$.
Therefore, to prove {\it Theorem}~1, it is equivalent to prove that $Y_N\rightarrow \mathbb{E}[Y_N]$ when $N$ is large.
However, since $x[\tilde n]e^{j\varphi_{\tilde n}}, \forall n$ are not i.i.d. complex random variables (RVs), we cannot directly use the conventional law of large numbers to give the proof.
Here, we consider the Chebyshev's inequality
\begin{equation}\label{Chb}
\Pr\left(\left|Y_N-\mathbb{E}\left[Y_N\right]\right|\ge\epsilon\right)\le \frac{\mathrm{Var}\left(Y_N\right)}{\epsilon^2}, \forall \epsilon > 0.
\end{equation}
Therefore, to prove that $Y_N\rightarrow \mathbb{E}[Y_N]$ when $N$ is large, we only need to prove that ${\mathrm{Var}(Y_N)}/{\epsilon^2}\rightarrow 0$, as $N$ goes to large.
The variance of $Y_N$ is given by
\begin{equation}\label{vY_N}
\begin{aligned}
&\mathrm{Var}\left(Y_N\right) = \mathrm{Var}\left(\frac{1}{N}\sum\nolimits_{\tilde{n}=1}^Nx[\tilde{n}]e^{j\varphi_{\tilde{n}}}\right)
\\
&=\frac{1}{N^2}\sum\limits_{\tilde{n}_1=1}^N\sum\limits_{\tilde{n}_2=1}^Ne^{j(\varphi_{\tilde{n}_1}-\varphi_{\tilde{n}_2})}\mathrm{Cov}\left(x[\tilde{n}_1],x[\tilde{n}_2]\right),
\end{aligned}
\end{equation}
where $\mathrm{Cov}\left(x[\tilde{n}_1],x[\tilde{n}_2]\right)$ denotes the covariance between $x[\tilde{n}_1]$ and $x[\tilde{n}_2]$, which is given by
\begin{equation}
\mathrm{Cov}\left(x[\tilde{n}_1],x[\tilde{n}_2]\right)=\mathbb{E}\left[x[\tilde{n}_1]x^\dagger[\tilde{n}_2]\right]-\mathbb{E}\left[x[\tilde{n}_1]\right]\mathbb{E}\left[x^\dagger[\tilde{n}_2]\right].
\end{equation}
Note that as the statistical properties of $x[\tilde{n}]$ are dependent on whether $\kappa_i+\tau=\kappa_j+\tau_p$, i.e., $\kappa_i-\kappa_j=\tau_p-\tau\triangleq d_\tau$, or not, we have the following two cases:

{\it \textbf{Case-I}}: For $\kappa_i+\tau=\kappa_j+\tau_p$, i.e., $\kappa_i-\kappa_j=d_\tau$, we have
\begin{equation}
\mathbb{E}\left[x[\tilde{n}]\right] = \mathbb{E}\left[\left|s[\tilde{n}-\kappa_i-\tau]\right|^2\right].
\end{equation}
Note that since the information-bearing symbols $s[\tilde n],\forall n$ are i.i.d. complex RVs, satisfying $s[\tilde{n}]\overset{i.i.d.}{\sim}\mathcal{CN}(0,1)$, we have $\mathbb{E}\left[x[\tilde{n}]\right]=1$.
By substituting it into \eqref{eY_N}, we have
\begin{equation}\label{eY_N1}
\mathbb{E}\left[Y_N\right]=\frac{1}{N}\sum\nolimits_{{\tilde{n}}=1}^Ne^{j\varphi_{\tilde{n}}} = \frac{1}{N}\sum\nolimits_{{\tilde{n}}=1}^Ne^{j2\pi d_\nu {\tilde{n}}T_s}\triangleq \psi(d_\nu ).
\end{equation}
Moreover, as $s[\tilde{n}_1]$ and $s[\tilde{n}_2],\forall n_1\neq n_2$ are i.i.d. complex RVs, we have $\mathbb{E}\left[x[\tilde{n}_1]x^\dagger[\tilde{n}_2]\right]=\mathbb{E}\left[x[\tilde{n}_1]\right]\mathbb{E}\left[x^\dagger[\tilde{n}_2]\right]$. Therefore, the covariance between $x[\tilde{n}_1]$ and $x[\tilde{n}_2]$ is $\mathrm{Cov}\left(x[\tilde{n}_1],x[\tilde{n}_2]\right)=0$, $\forall n_1\neq n_2$,  and \eqref{vY_N} reduces to
\begin{equation}\label{vY_N2}
\mathrm{Var}(Y_N)=\frac{1}{N^2}\sum\limits_{\tilde{n}=1}^N\mathrm{Var}(x[\tilde{n}])=\frac{\xi_1}{N},
\end{equation}
where $\xi_1\triangleq\mathrm{Var}(x[\tilde{n}])$ is a positive constant, denoting the variance of $x[\tilde{n}]$.
Therefore, by substituting \eqref{vY_N2} into \eqref{Chb}, when $N$ is large, we have
\begin{equation}\label{Chb2}
\Pr\left(\left|Y_N-\mathbb{E}\left[Y_N\right]\right|\ge\epsilon\right)\le \frac{\xi_1}{N\epsilon^2}\rightarrow 0, \forall \epsilon > 0.
\end{equation}
Thus the proof of {\it Theorem} 1 for $\kappa_i-\kappa_j=d_\tau$ is completed.

{\it \textbf{Case-II}}: For $\kappa_i+\tau\neq \kappa_j+\tau_p$, i.e., $\kappa_i-\kappa_j \neq d_\tau$, we have
\begin{equation}\label{eX_n}
\begin{aligned}
\mathbb{E}\left[x[\tilde{n}]\right]&=\mathbb{E}\left[s[\tilde{n}-\kappa_i-\tau]s^\dagger[\tilde{n}-\kappa_j-\tau_p]\right]\\
&=\mathbb{E}\left[s[\tilde{n}-\kappa_i-\tau]\right]\mathbb{E}\left[s^\dagger[\tilde{n}-\kappa_j-\tau_p]\right]=0.
\end{aligned}
\end{equation}
Thus, by substituting \eqref{eX_n} into \eqref{eY_N}, we have
\begin{equation}\label{eY_N2}
\mathbb{E}\left[Y_N\right] = \frac{1}{N}\sum\limits_{\tilde{n}=1}^N\mathbb{E}\left[x[\tilde{n}]\right]e^{j\varphi_{\tilde{n}}}=0.
\end{equation}
Furthermore, $\forall {\tilde{n}}_1\neq {\tilde{n}}_2$, we have
\begin{equation}\label{x_v_12}
\begin{aligned}
\mathbb{E}\left[x[{\tilde{n}}_1]x^\dagger[{\tilde{n}}_2]\right] &= \mathbb{E}\big[s[{\tilde{n}}_1-\kappa_i-\tau]s^\dagger[{\tilde{n}}_1-\kappa_j-\tau_p]\\
&\quad\times s^\dagger[{\tilde{n}}_2-\kappa_i-\tau]s[{\tilde{n}}_2-\kappa_j-\tau_p]\big],
\end{aligned}
\end{equation}
where depending on whether ${\tilde{n}}_1-\kappa_i-\tau={\tilde{n}}_2-\kappa_j-\tau_p$ and/or ${\tilde{n}}_1-\kappa_j-\tau_p={\tilde{n}}_2-\kappa_i-\tau$ or not, we have the following subcases:

1): if ${\tilde{n}}_1-\kappa_i-\tau={\tilde{n}}_2-\kappa_j-\tau_p$ but ${\tilde{n}}_1-\kappa_j-\tau_p\neq {\tilde{n}}_2-\kappa_i-\tau$, \eqref{x_v_12} reduces to
\begin{equation}\label{x_v_nq}
\begin{aligned}
\mathbb{E}\left[x[{\tilde{n}}_1]x^\dagger[{\tilde{n}}_2]\right] &= \mathbb{E}\left[(s[{\tilde{n}}_1-\kappa_i-\tau])^2\right]\mathbb{E}\left[s^\dagger[{\tilde{n}}_1-\kappa_j-\tau_p]\right]
\\
&\quad\times\mathbb{E}\left[s^\dagger[{\tilde{n}}_2-\kappa_i-\tau]\right] = 0.
\end{aligned}
\end{equation}

2): if ${\tilde{n}}_1-\kappa_i-\tau\neq {\tilde{n}}_2-\kappa_j-\tau_p$ but ${\tilde{n}}_1-\kappa_j-\tau_p =  {\tilde{n}}_2-\kappa_i-\tau$, \eqref{x_v_12} reduces to
\begin{equation}
\begin{aligned}
\mathbb{E}\left[x[{\tilde{n}}_1]x^\dagger[{\tilde{n}}_2]\right] &= \mathbb{E}\left[(s^\dagger[{\tilde{n}}_1-\kappa_j-\tau_p)^2]\right]\mathbb{E}\left[s[{\tilde{n}}_1-\kappa_i-\tau]\right]
\\
&\quad\times\mathbb{E}\left[s[{\tilde{n}}_2-\kappa_j-\tau_p]\right] = 0.
\end{aligned}
\end{equation}

3): if ${\tilde{n}}_1-\kappa_i-\tau\neq {\tilde{n}}_2-\kappa_j-\tau_p$ and ${\tilde{n}}_1-\kappa_j-\tau_p \neq  {\tilde{n}}_2-\kappa_i-\tau$, \eqref{x_v_12} reduces to
\begin{equation}\label{x_v_ne2}
\begin{aligned}
&\mathbb{E}\left[x[{\tilde{n}}_1]x^\dagger[{\tilde{n}}_2]\right] = \mathbb{E}\left[s[{\tilde{n}}_1-\kappa_i-\tau]\right]\mathbb{E}\left[s^\dagger[{\tilde{n}}_1-\kappa_j-\tau_p]\right]
\\
&\qquad\times\mathbb{E}\left[s^\dagger[{\tilde{n}}_2-\kappa_i-\tau]\right]\mathbb{E}\left[s[{\tilde{n}}_2-\kappa_j-\tau_p]\right] = 0.
\end{aligned}
\end{equation}
Note that the two conditions ${\tilde{n}}_1-\kappa_i-\tau={\tilde{n}}_2-\kappa_j-\tau_p$ and ${\tilde{n}}_1-\kappa_j-\tau_p= {\tilde{n}}_2-\kappa_i-\tau$ cannot hold at the same time,
since $\kappa_i+\tau\neq\kappa_j+\tau_p$ and $n_1\neq n_2$.
Therefore, from \eqref{x_v_nq} to \eqref{x_v_ne2},  we have $\mathbb{E}\left[x[{\tilde{n}}_1]x^\dagger[{\tilde{n}}_2]\right] =0$, $\forall {\tilde{n}}_1\neq {\tilde{n}}_2$.
Moreover, since $\mathbb{E}\left[x[\tilde{n}]\right]=0, \forall \tilde{n}$ as in \eqref{eX_n}, we have $\mathbb{E}\left[x[{\tilde{n}}_1]x^\dagger[{\tilde{n}}_2]\right]=\mathbb{E}\left[x[\tilde{n}_1]\right]\mathbb{E}\left[x^\dagger[\tilde{n}_2]\right]=0$.
Therefore, the covariance between $x[\tilde{n}_1]$ and $x[\tilde{n}_2]$ is
$\mathrm{Cov}\left(x[\tilde{n}_1],x[\tilde{n}_2]\right)=0$, $\forall n_1\neq n_2$,  and \eqref{vY_N} reduces to
\begin{equation}\label{vY_N3}
\mathrm{Var}(Y_N)=\frac{1}{N^2}\sum\limits_{\tilde{n}=1}^N\mathrm{Var}(x[\tilde{n}])=\frac{\xi_2}{N},
\end{equation}
where $\xi_2\triangleq\mathrm{Var}(x[\tilde{n}])$ is a positive constant, denoting the variance of $x[\tilde{n}]$ for Case-II.
Therefore, similar to \eqref{Chb2}, when $N$ is large, we have
\begin{equation}\label{Chb3}
\Pr\left(\left|Y_N-\mathbb{E}\left[Y_N\right]\right|\ge\epsilon\right)\le \frac{\xi_2}{N\epsilon^2}\rightarrow 0, \forall \epsilon > 0.
\end{equation}
Thus the proof of {\it Theorem} 1 for $\kappa_i-\kappa_j\neq d_\tau$ is completed.

\section{Proof of Proposition 1}\label{appendix b}
The objective function of $\mathrm{P1.1}$ in \eqref{P1.1} can be written as
\begin{equation}
\left|\sum\nolimits_{l=1}^L\mathbf{h}_l^H\mathbf{f}_l\right|^2 = \left|\mathrm{Tr}\left(\mathbf{H}^H\mathbf{F}\right)\right|^2,
\end{equation}
where $\mathrm{Tr}(\cdot)$ denotes the trace of a square matrix.
Since $\mathrm{Tr}(\mathbf{H}^H\mathbf{F})=\mathrm{vec}(\mathbf{H})^H\mathrm{vec}(\mathbf{F})$, \eqref{P1.1} can be further written as
\begin{equation}
\left|\sum\nolimits_{l=1}^L\mathbf{h}_l^H\mathbf{f}_l\right|^2 = \mathrm{vec}(\mathbf{F})^H\mathrm{vec}(\mathbf{H})\mathrm{vec}(\mathbf{H})^H\mathrm{vec}(\mathbf{F}),
\end{equation}
Similarly, we have
\begin{equation}
\sum\nolimits_{l=1}^L\left\|\mathbf{f}_l\right\|^2 = \mathrm{Tr}\left(\mathbf{F}^H\mathbf{F}\right) = \mathrm{vec}(\mathbf{F})^H\mathrm{vec}(\mathbf{F}),
\end{equation}
and
\begin{equation}
\begin{aligned}
&{{\mathbf{a}^H(\theta)\left(\sum\nolimits_{l=1}^L\mathbf{f}_l\mathbf{f}_l^H\right)\mathbf{a}(\theta)}} = \mathrm{Tr}\left(\mathbf{F}^H\mathbf{A}(\theta)\mathbf{F}\right)\\
&=\mathrm{vec}(\mathbf{F})^H\mathrm{vec}(\mathbf{A}(\theta)\mathbf{F})=\mathrm{vec}(\mathbf{F})^H\left(\mathbf{I}_L\otimes\mathbf{A}(\theta)\right)\mathrm{vec}(\mathbf{F}).
\end{aligned}
\end{equation}
For the LHS of the ISR constraint in \eqref{ISR_cons2}, according to \eqref{AF_mis}, it can be first written as
\begin{equation}
\begin{aligned}
&{\sum\limits_{0<\left|d_\tau\right|\le n_d}\left|\mathbf{a}^H(\theta)\left(\sum\nolimits_{(i,j)\in\mathcal{S}(d_\tau)}\mathbf{f}_i\mathbf{f}_j^H\right)\mathbf{a}(\theta)\right|^2} \\
&= \sum\limits_{0<|d_\tau|\le n_d}\left|\mathbf{a}^H(\theta)\mathbf{F}\bm{\Lambda}(d_\tau,0)\mathbf{F}^H\mathbf{a}(\theta)\right|^2,
\end{aligned}
\end{equation}
which can be further written as
\begin{equation}
\begin{aligned}
&\sum\limits_{0<|d_\tau|\le n_d}\left|\mathbf{a}^H(\theta)\mathbf{F}\bm{\Lambda}(d_\tau,0)\mathbf{F}^H\mathbf{a}(\theta)\right|^2 \\
&= \sum\limits_{0<|d_\tau|\le n_d}\left|\mathrm{Tr}\left(\bm{\Lambda}(d_\tau,0)\mathbf{F}^H\mathbf{A}(\theta)\mathbf{F}\right)\right|^2 \\
&= \sum\limits_{0<|d_\tau|\le n_d}\left|\mathrm{vec}\left(\mathbf{A}^H(\theta)\mathbf{F}\bm{\Lambda}^T(d_\tau,0)\right)^H\mathrm{vec}(\mathbf{F})\right|^2\\
&=\sum\limits_{0<|d_\tau|\le n_d}\left|\left(\left(\bm{\Lambda}(d_\tau,0)\otimes\mathbf{A}^H(\theta)
\right)\mathrm{vec}(\mathbf{F})\right)^H\mathrm{vec}(\mathbf{F})\right|^2\\
& = {\sum\limits_{0<\left|d_\tau\right|\le n_d}\left|\mathrm{vec}(\mathbf{F})^H\left(\bm{\Lambda}^T(d_\tau,0)\otimes\mathbf{A}(\theta)\right)\mathrm{vec}(\mathbf{F})\right|^2}.
\end{aligned}
\end{equation}
Therefore, the problem $\mathrm{(P1.1)}$ can be rewritten as $\mathrm{(P1.2)}$ accordingly.
\end{appendices}
\bibliographystyle{IEEEtran}
\bibliography{DAM}

\end{document}